\documentclass[aps,pra,10pt,twocolumn,superscriptaddress]{revtex4-1}

\usepackage{amsmath}    % need for subequations
\usepackage{graphicx}   % need for figures
\usepackage{verbatim}   % useful for program listings
\usepackage{color}      % use if color is used in text
\usepackage{subfigure}	% use for side-by-side figures
\usepackage{hyperref}   % use for hypertext links, including those to external documents and URLs
\usepackage{mathtools}

\newcommand{\lef}{\left(}
\newcommand{\rig}{\right)}

\newcommand{\imu}{\mathrm i}

\newcommand{\deriv}{\mathrm d}

\newcommand{\e}{\mathrm e}
\newcommand{\kB}{k_\mathrm{B}}

\newcommand{\EF}{E_\mathrm{F}}
\newcommand{\kF}{k_\mathrm{F}}
\newcommand{\vF}{v_\mathrm{F}}

\newcommand{\veck}{\mathbf k}
\newcommand{\vecv}{\mathbf v}
\newcommand{\vecvF}{\mathbf{v}_\mathrm{F}}

\newcommand{\vecE}{\mathbf E}

\newcommand{\eq}{\mathrm{eq}}
\newcommand{\GB}{\mathrm{GB}}
\newcommand{\BS}{\mathrm{BS}}
\newcommand{\bulk}{\mathrm{bulk}}
\newcommand{\EMA}{\mathrm{EMA}}

\begin{document}

\title{Resistivity scaling model for metals with conduction band anisotropy}

\author{Miguel De Clercq}
\affiliation{Physics Department, University of Antwerp, Groenenborgerlaan 171, B-2020 Antwerpen, Belgium}
\author{Kristof Moors}
\email[E-mail: ]{kristof.moors@uni.lu}
\affiliation{Physics and Materials Science Research Unit, University of Luxembourg, Avenue de la Fa\"iencerie 162a, L-1511 Luxembourg, Luxembourg}
\author{Kiroubanand Sankaran}
\affiliation{imec, Kapeldreef 75, B-3001 Leuven, Belgium}
\author{Geoffrey Pourtois}
\affiliation{imec, Kapeldreef 75, B-3001 Leuven, Belgium}
\author{Shibesh Dutta}
\affiliation{imec, Kapeldreef 75, B-3001 Leuven, Belgium}
\affiliation{Department of Physics and Astronomy, KU Leuven, Celestijnenlaan 200D, B-3001 Leuven, Belgium}
\author{Christoph Adelmann}
\affiliation{imec, Kapeldreef 75, B-3001 Leuven, Belgium}
\author{Wim Magnus}
\affiliation{imec, Kapeldreef 75, B-3001 Leuven, Belgium}
\affiliation{Physics Department, University of Antwerp, Groenenborgerlaan 171, B-2020 Antwerpen, Belgium}
\author{Bart Sor\'ee}
\affiliation{imec, Kapeldreef 75, B-3001 Leuven, Belgium}
\affiliation{Physics Department, University of Antwerp, Groenenborgerlaan 171, B-2020 Antwerpen, Belgium}
\affiliation{Electrical Engineering (ESAT) Department, KU Leuven, Kasteelpark Arenberg 10, B-3001 Leuven, Belgium}

\date{\today}

\begin{abstract}
It is generally understood that the resistivity of metal thin films scales with film thickness mainly due to grain boundary and boundary surface scattering. Recently, several experiments and \textit{ab initio} simulations have demonstrated the impact of crystal orientation on resistivity scaling. The crystal orientation cannot be captured by the commonly used resistivity scaling models and a qualitative understanding of its impact is currently lacking. In this work, we derive a resistivity scaling model that captures grain boundary and boundary surface scattering as well as the anisotropy of the band structure. The model is applied to Cu and Ru thin films, whose conduction bands are (quasi-) isotropic and anisotropic respectively. After calibrating the anisotropy with \textit{ab initio} simulations, the resistivity scaling models are compared to experimental resistivity data and a renormalization of the fitted grain boundary reflection coefficient can be identified for textured Ru.
\end{abstract}

\maketitle

\section{Introduction} \label{sec:introduction}
Many semiclassical and quantum mechanical resistivity scaling models (see \cite{fuchs1938conductivity, sondheimer1952mean,mayadas1970electrical,soffer1967statistical,
trivedi1988quantum,fishman1989surface,zhang1995conductivity,
palasantzas1998surface,meyerovich2002surface,Rickman2012,Moors2016,Munoz2017})
have been developed over the last decades and have provided a satisfactory description for the thickness dependent resistivity of metal thin films down to nanometer scale thicknesses \cite{de1988temperature,liu2001thickness,
tay2005electrical,camacho2006surface,sun2010surface,chawla2011electron}.
The thickness dependence can mainly be attributed to grain boundary and boundary surface scattering, whose impact on the resistivity increases when the film thickness is reduced. For grain boundaries, this is a consequence of the typical observation that grain sizes in polycrystalline films decrease with decreasing thickness.

Fuchs and Sondheimer developed a seminal semiclassical model that describes the impact of thin film boundary surface scattering on the resistivity by applying diffuse or partially diffuse boundary conditions \cite{fuchs1938conductivity,sondheimer1952mean}. Mayadas and Shatzkes later included the impact of grain boundary scattering without invoking Matthiessen's rule \cite{mayadas1970electrical}. The resulting expression is still widely used today as it provides an analytical expression of the thin film resistivity as a function of its thickness, allowing for a straightforward analysis of experimental resistivity data and determining the relative impact of grain and film boundaries. The model is typically being considered with two fitting parameters that respectively represent the average reflection coefficient of the grain boundaries and the specularity of the thin film boundary surfaces (due to e.g. atomic-scale boundary roughness). In this way, the relative contribution of grain boundary and boundary surface scattering for the resistivity degradation can easily be read from the fitting parameters.

The Mayadas-Shatzkes model shines through its simplicity and wide applicability, but it is derived within the framework of the simplest effective mass description for the conduction bands. This offers a reasonable description in case the metal is nearly isotropic. However, it is not necessarily adequate for many metals in the periodic table whose band structure deviates significantly from an isotropic band structure, e.g. exhibiting an anisotropic Fermi velocity and multiple bands with multiple electron or hole pockets centered around different symmetry points in the Brillouin zone (e.g. Co, W, Os, Ru, Ir) \cite{sankaran2015exploring,gall2016electron}.
Particularly for textured thin films (see \cite{Dutta2017}) or nanowires that are grown along a specific crystal orientation \cite{choi2014failure,Jones2015,Hegde2016,Lanzillo2017}, one can doubt the validity of this approach.

Recently, Li \textit{et al.} proposed a phenomenological correction to the Fuchs-Sondheimer model for metal thin films of Os with a nonspherical Fermi surface to explain the experimental findings \cite{Li2017}. In this work, we derive a model to describe resistivity scaling of imperfect metal thin films, while retaining some features of the electronic structure to capture the impact of conduction band anisotropy. The model is similar in spirit to the Mayadas-Shatzkes model and provides the resistivity as a function of the film thickness in a straightforward manner with two fitting parameters representing grain boundary and film boundary surface scattering. The main extension of our model is the consideration of a diagonal effective mass tensor that is tailored for the metal and thin film texture under consideration.

The model is presented in section~\ref{sec:model} after briefly reviewing the Mayadas-Shatzkes model, followed by a subsection on the effective mass fitting procedure. Section~\ref{sec:discussion} contains a discussion of the results, implications for experiments and limitations of applicability of the model.
We conclude in section~\ref{sec:conclusion}.

\section{Model} \label{sec:model}
The conduction electrons of a thin film, with length $L$ along the transport direction, are modeled as quasi-free fermions residing in a single conduction band and being treated in the effective mass approximation. The film thickness is considered to be large enough such that the allowed three-dimensional wave vectors $\veck$, providing a unique (apart from the two-fold spin degeneracy) label for the different electron states, can safely be assumed to be quasi-continuous.

\subsection{Mayadas-Shatzkes model} \label{subsec:MS}
\begin{figure}[tb]
	\centering
	\includegraphics[width=0.8\linewidth]{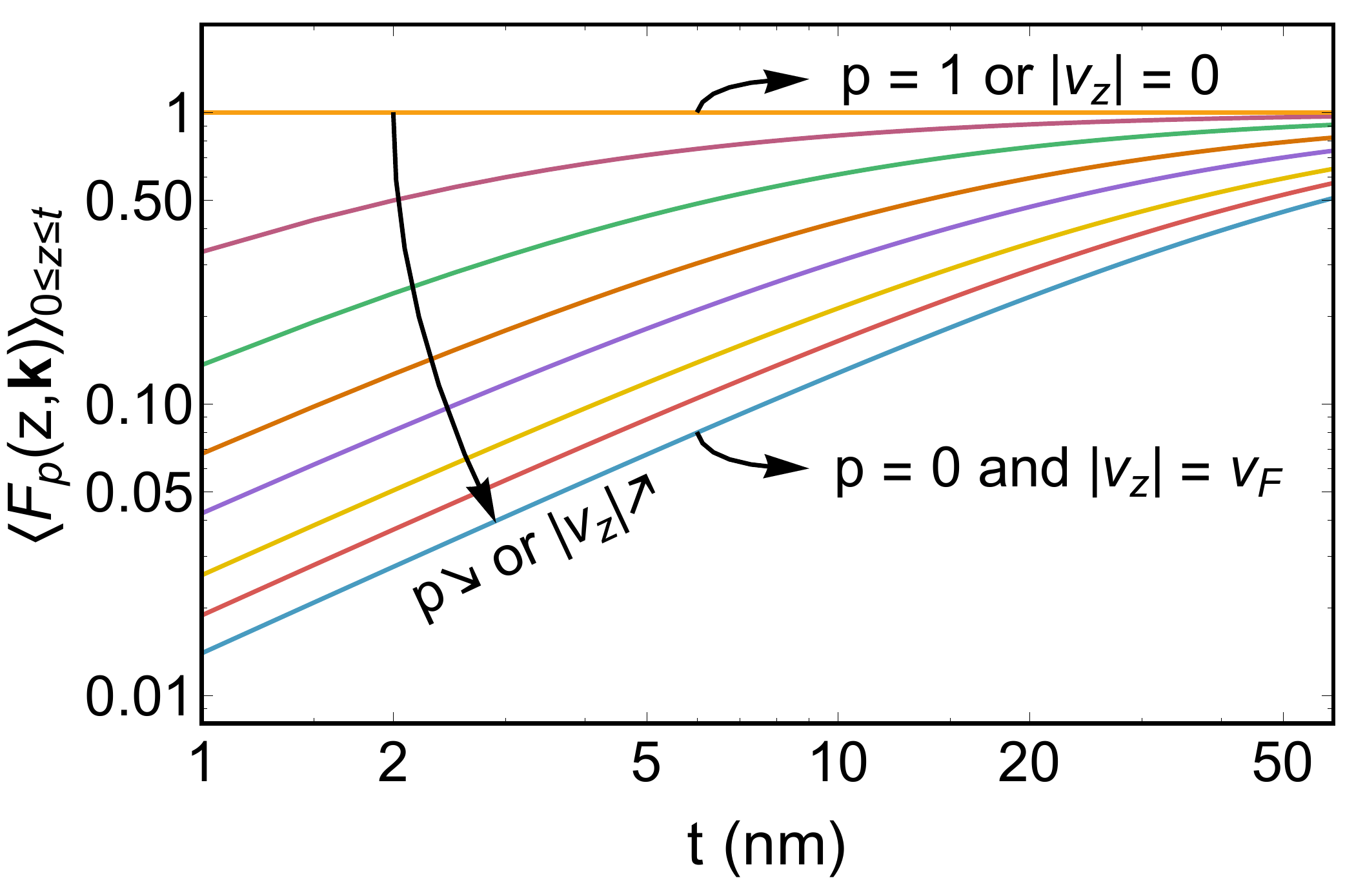}
	\caption{
		The Fuchs fraction $F_p(z, \veck)$, defined in Eq.~(\ref{eq:BS_boundary_cond}), is evaluated as a function of the film thickness $t$, showing its average over the position normal to the film boundaries for the full range of specularity parameters $p$ ($0 \leq p \leq 1$) and transverse velocities $v_z$ ($-\vF \leq v_z \leq \vF$). The bulk limit is retrieved (Fuchs fraction equal to one) when $p=1$ or $v_z=0$ and the lower bound when $p=0$ and $|v_z| = \vF$.
    The results are obtained for $\tau^*(\veck) = 25$~fs and $\vF = 1.57 \cdot 10^6$~m/s, resembling a monocrystalline Cu thin film.
		}
\label{fig:Fuchs}
\end{figure}

Mayadas and Shatzkes proposed a model for the resistivity scaling of polycrystalline films with reduced thicknesses due to an increase of grain boundary and boundary surface scattering \cite{mayadas1970electrical}. The grain boundaries are modeled by a sequence of delta-function potential barriers normal to the transport direction $x$, which can be regarded as effective representations for an ensemble average of grain boundaries with different shapes and orientations:
\begin{equation} \label{eq:GB_Pot}
  V^\GB \equiv \sum_{i=1}^N S \, \delta(x - x_i).
\end{equation}
$N$ such barriers are being considered, leading to an average grain boundary separation $d = L / N$, which can be understood as the mean linear intercept for a random straight trajectory through a thin film sample along the transport direction. This can easily be extracted from plan-view TEM images for example \cite{standard1996e112,Dutta2017}. The barrier positions $x_i$ are considered to be distributed according to a Gaussian distribution function $g(x_1, \ldots, x_N)$,
\begin{equation} \label{eq:Gaussian}
  g(x_1, \ldots, x_N) \equiv \frac{1}{L} \prod\limits_{i=1}^{N-1}
    \frac{\exp\left[
      -(x_{i+1} - x_i - d)^2 / (2 s^2) \right]}{(2 \pi s^2)^{1/2}},
\end{equation}
with standard deviation $s$ for the mean linear intercept. The Boltzmann transport equation is then considered to compute the distribution function $f(\veck)$. Keeping only the lowest-order contributions and assuming a small constant electric field vector $\vecE$ oriented along the transport direction, the stationary Boltzmann equation reduces to
\begin{equation} \label{eq:BTE} \begin{split}
  &e E \, v_x(\veck) \frac{\partial f^\eq(\epsilon(\veck))}{\partial \epsilon} \\
  &\; = \sum_{\veck'} P(\veck, \veck') \left[ f_1(\veck') - f_1(\veck) \right] - \frac{f_1(\veck)}{\tau(\veck)},
\end{split} \end{equation}
where $e$ is the electron charge, $\epsilon(\veck)$ the electron state energy, $v_x(\veck)$ the $x$-component of the electron velocity, $\tau(\veck)$ the bulk collision time due to impurities, defects and electron-phonon interactions, $P(\veck, \veck')$ the scattering rate to go from $\veck$ to $\veck'$ (or the opposite) as a result of elastic grain boundary scattering and $f_1(\veck) \equiv f(\veck) - f^\eq(\epsilon(\veck))$ with $f^\eq(\epsilon(\veck))$ the Fermi-Dirac distribution. Scattering at boundary surfaces is not yet included in this equation. The scattering rates are calculated with Fermi's golden rule and an averaging over the distribution $g(x_1, \ldots, x_N)$, leading to
\begin{equation} \label{eq:GB_Scat} \begin{split}
  P(\veck, \veck') &= F(|k_x|) \,
    \delta_{\veck_\perp, \veck_\perp'} \, \delta_{k_x, -k_x'}, \\
  F(|k_x|) &\equiv \frac{\hbar \kF^2}{m_\e |k_x| d} \frac{R}{1 - R} \\
    &\mkern-10mu \times \frac{1 - \exp(-4 k_x^2 s^2)}{1 + \exp(-4 k_x^2) -
      2\exp(-k_x^2 s^2) \cos(2 k_x d)}, \\
  R &= 1 \left/ \left[ 1 + \hbar^4 \kF^2/(m_\e S)^2 \right] \right.,
  \quad \veck_\perp \equiv (k_y, k_z) .
\end{split} \end{equation}
The derivation can be found in Appendix~\ref{sec:appendix1}. The reflection coefficient $R$ for an electron at the Fermi level with wave vector perpendicularly oriented to a delta-function barrier was used to rewrite the transition probability. Note that the expression on the second line is corrected with a factor of two in comparison with Mayadas and Shatzkes. The solution of Eq.~(\ref{eq:BTE}) is then given by
\begin{equation} \label{eq:BTE_Sol} \begin{split}
  f_1(\veck) &= -\tau^*(\veck) \, e E \, v_x(\veck) \,
    \frac{\partial f^\eq(\epsilon(\veck))}{\partial \epsilon} , \\
  1/\tau^*(\veck) &\equiv 1/\tau(\veck) + 2 F(|k_x|).
\end{split} \end{equation}
For random grain boundary configurations of typical metallic thin films, it is safe to assume $\kF^2 s^2 \gg 1$, such that $F(|k_x|)$ reduces to
\begin{equation}
  F(|k_x|) = \frac{\hbar \kF^2}{m_\e |k_x| d} \frac{R}{1 - R}.
\end{equation}
The conductivity $\sigma_x^\GB$ (along transport direction $x$), taking into account the bulk scattering contribution and grain boundary scattering, can be calculated using \cite{jacoboni2010theory, mahan2013many}:
\begin{equation} \label{eq:BTE_Cond}
  \sigma_x^\GB = -2 e^2 \int \frac{\deriv^3 k}{(2 \pi)^3} \; v_x^2(\veck) \,
    \tau^*(\veck) \frac{\partial f^\eq(\epsilon(\veck))}{\partial \epsilon}.
\end{equation}
The result obtained by Mayadas and Shatzkes, assuming zero temperature (being a very reasonable assumption for typical metals at room temperature) and an isotropic bulk collision time $\tau$ (or equivalently an isotropic mean free path $l_0$), is given by
\begin{equation} \label{eq:GB_Cond} \begin{split}
  \sigma^\GB(\alpha) &= \frac{n_\e e^2 \tau}{m_\e}
    3 \left[ \frac{1}{3} - \frac{\alpha}{2} + \alpha^2 - \alpha^3
      \ln\lef \frac{1 + \alpha}{\alpha} \rig \right], \\
  n_\e &= \kF^3 / (3 \pi^2) =
      (2 m_\e \EF)^{3/2}/(3 \pi^2 \hbar^3), \\
  \alpha &\equiv 2\tau \frac{\hbar \kF}{m_\e d} \frac{R}{1 - R}
      = \frac{l_0}{d} \frac{2 R}{1 - R}, \quad
    l_0 \equiv \vF \tau,
\end{split} \end{equation}
with $n_\e$ the bulk electron density and $\vF$ the Fermi velocity: $\vF \equiv \hbar \kF / m_\e$.
Scattering at the thin film boundary surfaces is not taken into account through a scattering probability, but through boundary conditions on the Boltzmann distribution function, carrying an additional dependence on the coordinate $z$ normal to the film boundary surfaces. This approach was first introduced by Fuchs and Sondheimer without the inclusion of grain boundary scattering \cite{fuchs1938conductivity, sondheimer1952mean}. The Boltzmann equation becomes
\begin{equation} \label{eq:BTE_Surface} \begin{split}
  &v_z(\veck) \frac{\partial f_1(z, \veck)}{\partial z} + e E \, v_x(\veck)
    \frac{\partial f^\eq(\epsilon(\veck))}{\partial \epsilon}
    = -\frac{f_1(z, \veck)}{\tau^*(\veck)}, \\
  &f_1(0, k_x, k_y, k_z) = p f_1(0, k_x, k_y, -k_z) \quad (v_z(\veck) > 0), \\
  &\,f_1(t, k_x, k_y, k_z) = p f_1(t, k_x, k_y, -k_z) \quad \, (v_z(\veck) < 0),
\end{split} \end{equation}
where $t$ is the film thickness $p$ reflects the probability to scatter specularly at the boundary surfaces at $z=0$ and $z=t$, while diffuse scattering occurs with a probability $1 - p$. Note that the notion of a transverse velocity $v_z(\veck)$ is invalid for extremely thin films, where confinement heavily affects the band structure. Due to the large conduction electron density of typical metals, this region is limited to thin films with thicknesses up to a few atom layers, i.e. $\sim 1$~nm \cite{Dutta2017,moors2017resistivity}. The solution of the equation with surface scattering boundary conditions is given by
\begin{equation} \label{eq:BS_boundary_cond} \begin{split}
  f_1(z, \veck) &= - \tau^*(\veck) \, e E \, v_x(\veck)
   \frac{\partial f^\eq(\epsilon(\veck))}{\partial \epsilon} F_p(z, \veck), \\
  F_p(z, \veck) &\equiv 1 - \vartheta(k_z) \frac{(1 - p) \exp\left\{-z / \left[ \tau^*(\veck) v_z(\veck) \right] \right\}}{1 - p \exp\left\{-t / \left[ \tau^*(\veck) v_z(\veck) \right] \right\}} \\
  & - \vartheta(-k_z) \frac{(1 - p) \exp\left\{(t - z) / \left[ \tau^*(\veck) v_z(\veck) \right] \right\}}{1 - p \exp\left\{t / \left[ \tau^*(\veck) v_z(\veck) \right] \right\}},
\end{split} \end{equation}
with the Heaviside step function $\vartheta(k)$ and Fuchs fraction $F_p(z, \veck)$ (see Fig.~\ref{fig:Fuchs}). This solution leads to the following conductivity formula for thin films with grain boundary and partially diffusive boundary surface scattering on top of isotropic bulk collisions, $\sigma^{\GB + \BS}(\alpha, p)$:
\begin{widetext}
\begin{equation} \label{eq:cond_GB_BS}
  \sigma^{\GB + \BS}(\alpha, p) = \sigma^\GB(\alpha) -
    \frac{n_\e e^2 \tau}{m_\e}
    \frac{6}{\pi \kappa} (1 - p) \!
    \int\limits_0^{\pi/2} \! \deriv \theta \! \int\limits_0^{\pi/2} \! \deriv\phi \;
    \frac{\sin^3\!\theta \cos\theta \cos^2\!\phi}{H^2(\alpha, \theta, \phi)}
    \frac{1 - \exp \left[ - \kappa H(\alpha, \theta, \phi) / \cos\theta \right]}
      {1 - p \exp \left[ - \kappa H(\alpha, \theta, \phi) / \cos\theta \right]},
\end{equation}
\end{widetext}
with:
\begin{equation} \begin{split}
  \kappa &\equiv t / l_0, \\
    H(\alpha, \theta, \phi) &\equiv 1 + \alpha / |\sin\theta \, \cos\phi|,
\end{split} \end{equation}
and $\theta$ and $\phi$ respectively referring to the polar and azimuthal angle of the spherical coordinate system with poles located on the $z$-axis.

The bulk conductivity is retrieved as the limiting case corresponding to $R=0$ and $p=1$:
\begin{equation} \begin{split}
  \sigma^\bulk &= \sigma^{\GB + \BS} (\alpha = 0, p = 1) \\
  &= n_\e e^2 \tau / m_\e = n_\e e^2 l_0 / (m_\e \vF).
\end{split} \end{equation}
Also note that the conductivity is finite in the limit $p \rightarrow 0$ while:
\begin{equation} \begin{split}
  &\lim\limits_{\alpha \rightarrow +\infty} \sigma^{\GB + \BS} (\alpha, p) = \lim\limits_{R \rightarrow 1} \sigma^{\GB + \BS} (R, p) = 0, \\
  &\lim\limits_{\alpha \rightarrow +\infty} \sigma^\GB (\alpha) = \lim\limits_{R \rightarrow 1} \sigma^\GB (R) = 0, \\
  &\lim\limits_{p \rightarrow 1} \sigma^{\GB + \BS}(\alpha, p) = \sigma^\GB(\alpha).
\end{split} \end{equation}
Analogous limiting cases can be discovered in the extended model presented in the following section.

\subsection{Conduction band anisotropy} \label{subsec:extension}
We extend the Mayadas-Shatzkes model to account for an anisotropic conduction band, for which we introduce a diagonal effective mass tensor. The Fermi surface $\epsilon(\veck) = \EF$ is ellipsoidal and the following relations hold, while $\veck$ corresponds to any electron state at the Fermi level:
\begin{equation} \label{eq:anisotropicEM} \begin{split}
  \EF &= \frac{\hbar^2 k_x^2}{2 m_x} +  \frac{\hbar^2 k_y^2}{2 m_y} +
      \frac{\hbar^2 k_z^2}{2 m_z} = \frac{\hbar^2 \kF^2}{2 m_\e}, \\
  n_\e &= M_x M_y M_z \frac{\kF^3}{3 \pi^2}
      = M_x M_y M_z \frac{(2 m_\e \EF)^{3/2}}{3 \pi^2 \hbar^3}, \\
  \kF &\equiv \sqrt{2 m_\e \EF}/\hbar, \qquad
    M_{x,y,z} \equiv \sqrt{m_{x,y,z}/m_\e}.
\end{split} \end{equation}
For Eq.~(\ref{eq:GB_Cond}), an isotropic magnitude of the Fermi level velocities, mean free path and collision time are considered, which needs to be modified in case of conduction band anisotropy. The electron velocity of Fermi level states is now anisotropic and given by:
\begin{equation}
  |\vecv(\veck)| = \left|\nabla_{\veck} E(\veck) / \hbar \right| =
    \hbar \sqrt{\frac{k_x^2}{m_x^2} + \frac{k_y^2}{m_y^2} + \frac{k_z^2}{m_z^2}},
\end{equation}
and in general this velocity magnitude can be related to an anisotropic mean free path $l_0(\veck)$ and/or collision time $\tau(\veck)$ through $|\vecv(\veck)| = l_0(\veck) / \tau(\veck)$. The grain boundary scattering probability is still given by Eq.~(\ref{eq:GB_Scat}), but $R$, the reflection coefficient of a Fermi level electron with velocity perpendicular to a grain boundary barrier, is related differently to the barrier strength due to the band anisotropy:
\begin{equation} \label{eq:GB_Scat_Aniso}
  R = 1 \left/ \left[ 1 + \hbar^4 \kF^2/ (M_x m_\e S)^2 \right] \right..
\end{equation}
In order to compute the conductivity with Eq.~(\ref{eq:cond_GB_BS}), we need to integrate over an ellipsoidal Fermi surface. Hence, we define new integration variables to map it to a spherical surface integration:
\begin{equation} \label{eq:rescaling}
  k_{x, y, z} \rightarrow k_{x,y,z} / M_{x, y, z}.
\end{equation}
Following the derivation of Mayadas and Shatzkes, the conductivity will first be calculated taking only the scattering by grain boundaries into account. Scattering at the external surfaces will be accounted for subsequently. We will consider two cases in the subsections below, the first one assuming an isotropic collision time, the second one an isotropic mean free path. The former is relevant for electron-phonon interactions while the latter is more suitable for the low temperature regime dominated by impurity and defect scattering.

\subsubsection{Isotropic collision time}
Using the result of Eq.~(\ref{eq:GB_Scat_Aniso}) and the rescaling according to Eq.~(\ref{eq:rescaling}), we solve for the conductivity, given by Eq.~(\ref{eq:BTE_Cond}), for an ellipsoidal Fermi surface with isotropic collision time $\tau$:
\begin{equation} \begin{split}
  \sigma^\GB(\beta) &= \frac{n_\e e^2 \tau}{m_x}
    3 \left[ \frac{1}{3} - \frac{\beta}{2} + \beta^2 - \beta^3
    \ln\lef \frac{1 + \beta}{\beta} \rig \right], \\
  \beta &\equiv 2 \tau \, \frac{\hbar \kF}{M_x m_\e d} \frac{R}{1 - R}
    = \frac{\alpha}{M_x}.
\end{split} \end{equation}
In order to include scattering at the film boundary surfaces we adopt the Fuchs-Sondheimer approach as before:
\begin{widetext}
\begin{equation} \label{eq:cond_isoTau}
  \sigma^{\GB + \BS}(\beta, p) = \sigma^\GB(\beta) - \frac{n_\e e^2 \tau}{m_x}
    \frac{6}{\pi \lambda} (1 - p) \!
    \int\limits_0^{\pi/2} \! \deriv \theta \! \int\limits_0^{\pi/2} \! \deriv\phi \;
    \frac{\sin^3\!\theta \cos\theta \cos^2\!\phi}{H^2(\beta, \theta, \phi)}
    \frac{1 - \exp \left[ - \lambda H(\beta, \theta, \phi) / \cos\theta \right]}
      {1 - p \exp \left[ - \lambda H(\beta, \theta, \phi) / \cos\theta \right]},
\end{equation}
\end{widetext}
with:
\begin{equation} \begin{split}
  H(\beta, \theta, \phi) &\equiv 1 + \beta/|\sin\theta \cos\phi|, \\
   \lambda &\equiv t m_z/(\hbar M_z \kF \tau) = M_z \kappa.
\end{split} \end{equation}
The remaining integrations cannot be carried out analytically and are performed numerically. As a limiting case, the bulk conductivity equals
\begin{equation}
  \sigma^\bulk = \sigma^{\GB + \BS}(\beta = 0, p = 1) = n_\e e^2 \tau / m_x.
\end{equation}

\subsubsection{Isotropic mean free path}
In case of an isotropic mean free path and an ellipsoidal Fermi surface, the collision time is directional and given by
\begin{equation}
  \tau(\veck) = \frac{l_0 \hbar}{|\nabla_\veck E(\veck)|} = \frac{l_0}{\hbar \sqrt{k_x^2/m_x^2 + k_y^2/m_y^2 + k_z^2/m_z^2}}.
\end{equation}
The conductivity expression for grain boundary scattering cannot be obtained analytically in this case and is given by:
\begin{equation} \label{eq:res_GB_iso_MFP}
  \sigma^\GB(\alpha) = \frac{n_\e e^2 \tau_x}{m_x}
    \frac{6}{\pi} \! \int\limits_0^{\pi/2} \! \deriv \theta \!
    \int\limits_0^{\pi/2} \! \deriv \phi \; \frac{\sin^3\!\theta \cos^2\!\phi}
      {\zeta(\theta, \phi) + \alpha / |\sin\theta \cos\phi|},
\end{equation}
where $\tau_x \equiv l_0 m_x / (\hbar M_x \kF)$ and
\begin{equation}
  \zeta(\theta, \phi) \equiv
    \sqrt{\sin^2\!\theta \cos^2\!\phi + \sin^2\! \theta \sin^2\!\phi \,
      \frac{M_x^2}{M_y^2} + \cos^2\!\theta \, \frac{M_x^2}{M_z^2}} \, .
\end{equation}
The extension for scattering at the film boundary surfaces is again obtained in a straightforward manner:
\begin{widetext}
\begin{equation} \label{eq:cond_isoMFP}
  \sigma^{\GB + \BS}(\alpha, p) = \sigma^\GB(\alpha) -
    \frac{n_\e e^2 \tau_x}{m_x} \frac{6}{\pi \mu}
    (1 - p) \!
    \int\limits_0^{\pi/2} \! \deriv \theta \! \int\limits_0^{\pi/2} \! \deriv\phi \; \frac{\sin^3\!\theta \cos\theta \cos^2\!\phi}
      {G^2(\alpha, \theta, \phi)}
    \frac{1 - \exp \left[ - \mu G(\alpha, \theta, \phi) / \cos\theta \right]}
      {1 - p \exp \left[ - \mu G(\alpha, \theta, \phi) / \cos\theta \right]},
\end{equation}
\end{widetext}
with:
\begin{equation} \begin{split}
  G(\alpha, \theta, \phi) &\equiv \zeta(\theta, \phi) +
    \alpha/|\sin\theta \cos\phi|, \\
    \mu \equiv M_z t/(M_x l_0) &= \lambda/M_x = M_z \kappa / M_x.
\end{split} \end{equation}
The bulk limit ($\alpha = 0$, $p = 1$) also requires numerical integration in this case, with $1/\zeta(\theta, \phi)$ in the integrand.

\subsection{Directional effective mass fit}
\label{subsec:fitting_procedure}
The aim of the fitting procedure in this subsection is to obtain an anisotropic effective mass model with an ellipsoidal energy-momentum relation that captures the anisotropic bulk conductivity of an arbitrary Fermi surface in the best possible way, thereby providing appropriate directional effective mass values for the semiclassical resistivity scaling model. The computational procedure for extracting the conductivity is presented below for a particular transport direction (x) and without imposing any Fermi surface averaging (as might be appropriate for monocrystalline thin films). Nevertheless, it can be generalized to an arbitrary transport direction with any type of averaging (e.g. in-plane or isotropic) in a straightforward manner (see Appendix~\ref{sec:appendix2}).

We start from the bulk conductivity expression for an arbitrary Fermi surface as obtained from Eq.~(\ref{eq:BTE}), without grain boundary scattering term, and with an additional conduction band label $n$:
\begin{equation} \label{eq:cond_bulk_abinit}
  \sigma_x^\bulk = -\frac{e^2}{4\pi^3} \sum_n \int \! \deriv^3 v \;
      \frac{v_{x \, n}^2(\vecv) \tau_n(\vecv)}
        {|\partial \vecv / \partial \veck|} \,
        \frac{\partial f_n^\eq(\epsilon(\vecv))}{\partial \epsilon},
\end{equation}
Note that we consider an integral over group velocities rather than wave vectors as the directionality of the former is more physically relevant, underlying the semiclassical boundary conditions for partial specular surface scattering in Eq.~(\ref{eq:BS_boundary_cond}).
We can now introduce a directional bulk conductivity $\sigma_x^\bulk(\theta, \phi)$ such that the total bulk conductivity can be obtained by an integration over the unit sphere, representing all possible directions for the Fermi velocity:
\begin{equation}
  \sigma_x^\bulk = \frac{1}{4 \pi}
      \int \! \deriv^2 \Omega \; \sigma_x^\bulk(\theta, \phi),
\end{equation}
with $\sigma_x^\bulk(\theta, \phi)$ defined as:
\begin{equation} \begin{split}
  &\sigma_x^\bulk(\theta, \phi) \\
  &\, \equiv - \frac{e^2}{\pi^2} \! \int\limits_0^{+\infty} \! \!
    \deriv v \, v^2 \sum_n \frac{v_{x \, n}^2(\vecv) \tau_n(\vecv)}
        {\operatorname{det}|\partial \vecv(\veck) / \partial \veck|} \frac{\partial f_n^\eq(\epsilon(\vecv))}{\partial \epsilon}.
\end{split} \end{equation}
This quantity will be approximated by considering the Fermi surface of a single conduction band with ellipsoidal energy-momentum relation based on the effective mass approximation as introduced in Eq.~(\ref{eq:anisotropicEM}), which can be rewritten in terms of directional Fermi velocities $\vecvF \equiv (v_x, v_y, v_z)$ as
\begin{equation} \label{eq:ani_velocity}
  a_x v_x^2 + a_y v_y^2 + a_z v_z^2 = 1, \quad \!
    a_{x,y,z} \equiv M_{x,y,z}^2 m_\e/(2 \EF),
\end{equation}
or, similarly, as a function of the polar angle $\theta$, azimuthal angle $\phi$ and the magnitude of the directional Fermi velocity $\vF(\theta, \phi)$ as:
\begin{equation} \label{eq:energy_velocity}
  a_x^2 \sin^2\!\theta \cos^2\!\phi + a_y^2 \sin^2\!\theta \sin^2\!\phi + a_z^2 \cos^2\!\theta =1/\vF^2(\theta, \phi),
\end{equation}
such that we get an approximate bulk conductivity $\sigma_x^\EMA$:
\begin{equation}
  \sigma_x^\EMA = \frac{1}{4 \pi} \! \int \! \deriv^2 \Omega \;
    \sigma_x^\EMA(\theta, \phi),
\end{equation}
with approximated directional bulk conductivity:
\begin{equation}
  \sigma_x^\EMA(\theta, \phi) \equiv \frac{4 e^2 \EF^2}{\pi^2 \hbar^3} \, a_x a_y a_z \sin^2\!\theta \cos^2\!\phi \, \vF^5(\theta, \phi)
    \tau(\theta, \phi).
\end{equation}
We can now fit parameters $a_x$, $a_y$, $a_z$ and $\EF$ such that $\sigma_x^\bulk(\theta, \phi) \approx \sigma_x^\EMA(\theta, \phi)$ for all angles $\theta$ and $\phi$, properly reflecting the bulk conductivity contributions from the different velocity orientations. Note that this fitting procedure does not involve any matching of quantities involving electron states away from the Fermi level, such as the work function of the metal or the band curvature of the bottom of a conduction band emerging from the \textit{ab initio} band structure. Furthermore, by fitting in velocity space, the full distribution of electron states in the Brillouin zone (e.g. centered around different symmetry points) is not captured, as can be expected when adopting the effective mass approximation.
A suitable fit can be obtained by combining $\sigma_x^\bulk(\theta, \phi) \approx \sigma_x^\EMA(\theta, \phi)$ and Eq.~(\ref{eq:ani_velocity}) to obtain a system of equations which can be solved with the least-squares method. The resulting system of equations will depend on the functional form adopted for $\tau(\theta, \phi)$. In case of an isotropic bulk collision time, we obtain:
\begin{widetext}
\begin{equation} \label{eq:fit_iso_tau} \begin{split}
  &\forall \, \theta, \, \phi: \quad \tau(\theta, \phi) = \tau, \qquad
    \vF^2(\theta, \phi)
      \approx \left[\pi^2 \hbar^3 \sigma_x^\bulk(\theta, \phi) /
      ( 4 e^2 \EF^2 \, a_x a_y a_z \sin^2\!\theta \cos^2\!\phi \, \tau ) \right]^{2/5}, \\
  &\Rightarrow
    \frac{a_x}{(\EF^2 \, a_x a_y a_z)^{2/5}} \sin^2\!\theta \cos^2\!\phi +
    \frac{a_y}{(\EF^2 \, a_x a_y a_z)^{2/5}} \sin^2\!\theta \sin^2\!\phi +
    \frac{a_z}{(\EF^2 \, a_x a_y a_z)^{2/5}} \cos^2\!\theta =
    \lef \frac{4 \sin^2\!\theta \cos^2\!\phi \, e^2 \tau}
      {\pi^2 \hbar^3 \sigma_x^\bulk(\theta, \phi)} \rig^{2/5},
\end{split} \end{equation}
\end{widetext}
where the second line is obtained by plugging the result of the first line into Eq.~(\ref{eq:energy_velocity}) and bringing over factors such that the right-hand side is independent of the fitting parameters and the left-hand side contains three independent fitting parameters. In case of an isotropic mean free path $l_0$, we get in a completely analogous way:
\begin{widetext}
\begin{equation} \label{eq:fit_iso_MFP} \begin{split}
  &\forall \, \theta, \, \phi: \quad \tau(\theta, \phi) = l_0 / \vF(\theta, \phi), \qquad \vF^2(\theta, \phi) \approx \left[ \pi^2 \hbar^3 \sigma_x^\bulk(\theta, \phi) /
      ( 4 e^2 \EF^2 \, a_x a_y a_z \, \sin^2\!\theta \cos^2\!\phi \, l_0 ) \right]^{1/2} \\
  &\Rightarrow
    \frac{a_x}{\EF (a_x a_y a_z)^{1/2}} \sin^2\!\theta \cos^2\!\phi +
    \frac{a_y}{\EF (a_x a_y a_z)^{1/2}} \sin^2\!\theta \sin^2\!\phi +
    \frac{a_z}{\EF (a_x a_y a_z)^{1/2}} \cos^2\!\theta = \lef \frac{4 \, \sin^2\!\theta \cos^2\!\phi \, e^2 l_0}{\pi^2 \hbar^3 \sigma_x^\bulk(\theta, \phi)} \rig^{1/2}.
\end{split} \end{equation}
\end{widetext}
As there are only three independent fitting parameters, parameters $a_x$, $a_y$, $a_z$ and $\EF$ cannot be uniquely determined. The fit is unaffected by a rescaling $\EF \rightarrow C \EF$, $M_{x,y,z} \rightarrow C^{-3/2} M_{x,y,z}$ when the collision time is isotropic,
and $\EF \rightarrow C \EF$, $M_{x,y,z} \rightarrow C^{-1/2} M_{x,y,z}$ when the mean free path is isotropic. This remaining degree of freedom can be eliminated by matching the density of states at the Fermi level of the anisotropic effective mass model and of the \textit{ab initio} band structure:
\begin{equation}
  (a_x a_y a_z)^{1/2} \frac{4 \EF^2}{\pi^2 \hbar^3} = -2 \sum_n \! \int \! \frac{\deriv^3 k}{(2 \pi)^3}
       \frac{\partial f_n^\eq(\epsilon(\veck))}{\partial \epsilon},
\end{equation}
with the density of states for the effective mass model on the left-hand side obtained from the electron density in Eq.~(\ref{eq:GB_Cond}) and the definition in Eq.~(\ref{eq:ani_velocity}). The Fermi level density of states of the anisotropic effective mass remains constant under rescaling $\EF \rightarrow C \EF$, $M_{x,y,z} \rightarrow C^{-1/6} M_{x,y,z}$, which differs from the invariance of the conductivity in the case of an isotropic collision time or mean free path.
For the numerical evaluation of $\sigma_x^\bulk(\theta, \phi)$, we have considered 500 pairs ($\theta_i$, $\phi_i$), uniformly distributed on the unit sphere, and approximated the derivative of the Fermi-Dirac distribution function at low temperatures ($\kB T \ll \EF$) by:
\begin{equation}
  \frac{\partial f_n^\eq(\epsilon(\veck))}{\partial \epsilon} \approx
    - \frac{\vartheta[2 \delta E - |\epsilon_n(\veck) - \EF|]}{4 \, \delta E},
\end{equation}
with $\vartheta(\epsilon)$ the Heaviside step function and $\delta E > 0$ very small.

\section{Results} \label{sec:results}
\begin{figure}[tb]
  \centering
  \subfigure[\ ]{\includegraphics[width=0.49\linewidth]{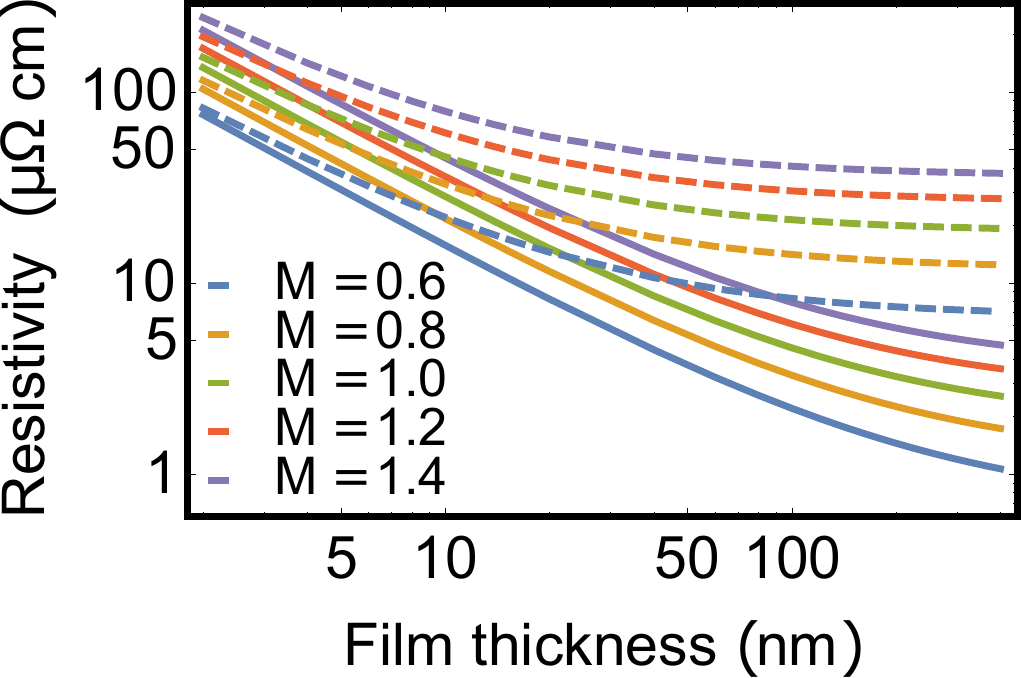}}
  \subfigure[\ ]{\includegraphics[width=0.49\linewidth]{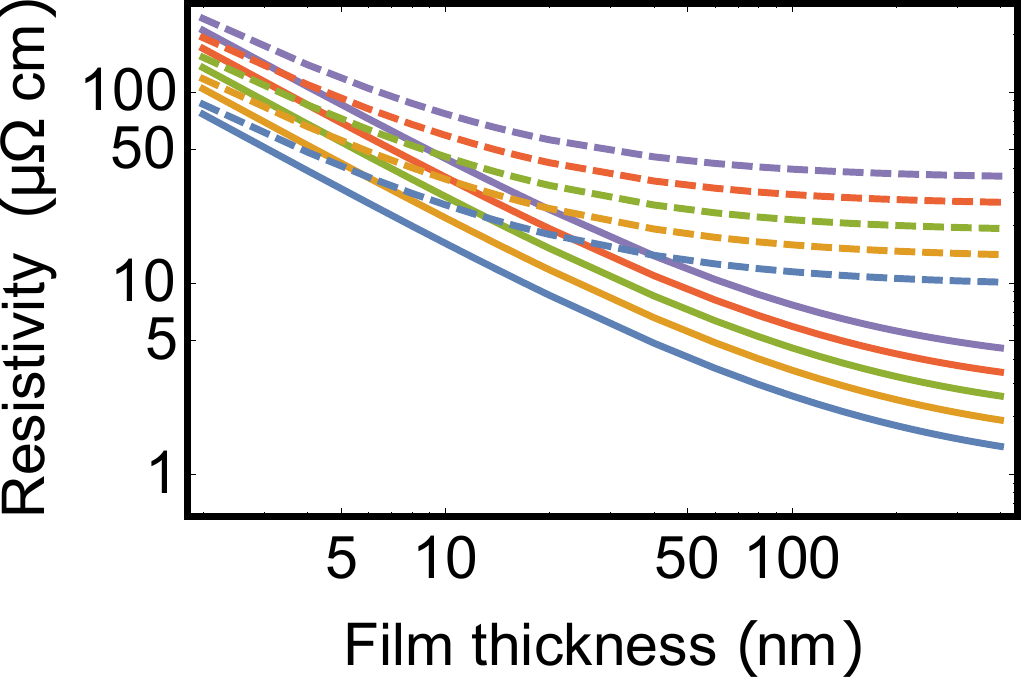}}
  \subfigure[\ ]{\includegraphics[width=0.49\linewidth]{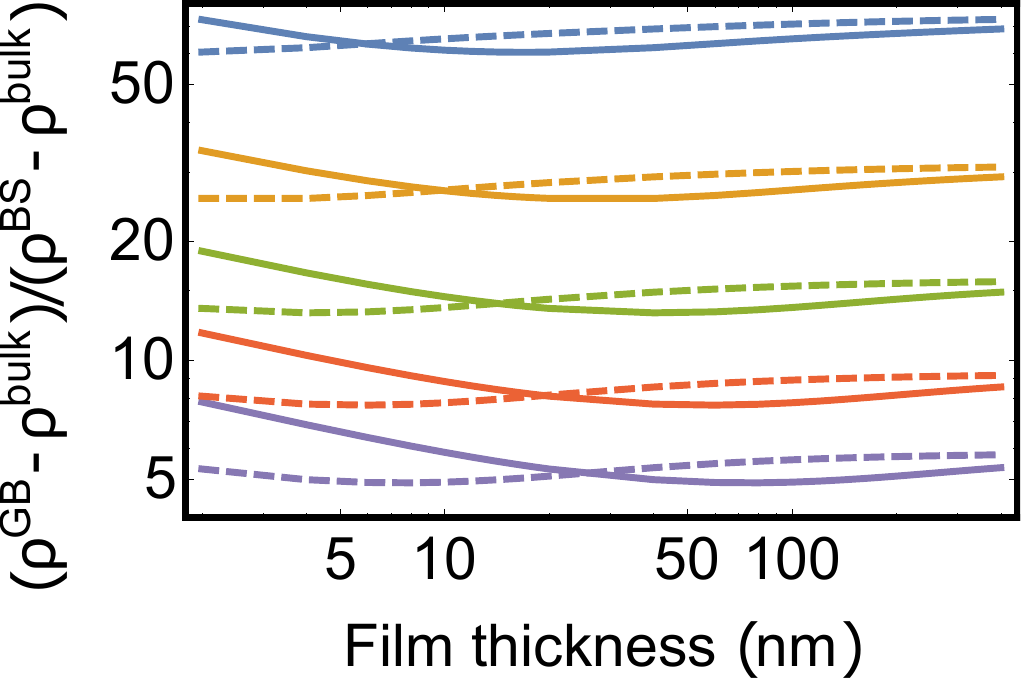}}
  \subfigure[\ ]{\includegraphics[width=0.49\linewidth]{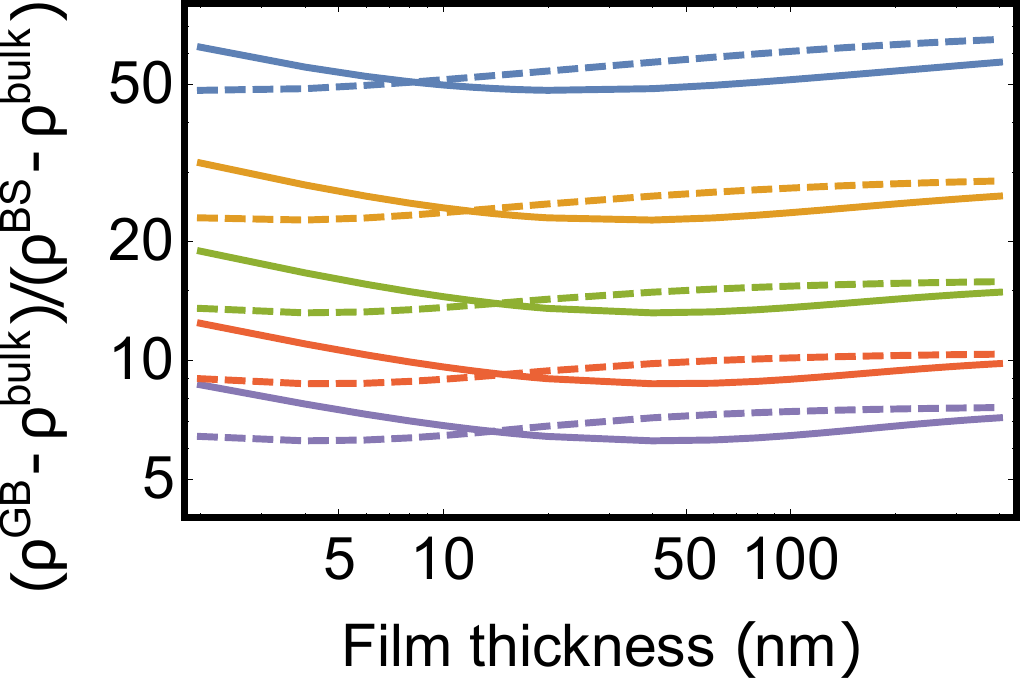}}
  \caption{
    (a-b) The resistivity is evaluated as a function of the film thickness, considering a Fermi energy of 5~eV ($n_\e \approx 50.8$~nm${}^{-3}$), reflection coefficient $R = 0.5$ and specularity parameter $p=0.5$ for different degrees of out-of-plane anisotropy by varying the eccentricity parameter $M$ ($M_x = M_y = M$, $M_z = 1/M^2$), assuming (a) an isotropic collision time of 37.7~fs (full lines) and 3.77~fs (dashed lines) (b) an isotropic mean free path of 50~nm (full lines) and 5~nm (dashed lines). The values of collision time and mean free path are equivalent choices when $M=1$.
    (c-d) The ratio of the resistivity contributions due to grain boundary and boundary surface scattering is evaluated as a function of the film thickness, considering an isotropic (c) collision time (d) mean free path.
  }
  \label{fig:res_scaling}
\end{figure}

The metal thin film conductivity (or resistivity) scaling formulas, derived under the assumption of an isotropic collision time (Eq.~(\ref{eq:cond_isoTau})) and an isotropic mean free path (Eq.~(\ref{eq:cond_isoMFP})), are evaluated for different parameters and degrees of conduction band anisotropy in Fig.~\ref{fig:res_scaling}, fixing the reflection coefficient and specularity parameter to the same value of 0.5 and the average linear intercept equal to the film thickness ($d = t$).
The following effective masses are considered: $M_x = M_y = M$ and $M_z = 1/M^2$, with eccentricity parameter $M > 0$. By varying $M$, the anisotropy can be controlled without changing the electron density nor the density of states at the Fermi level, easily verified by plugging these effective masses into the second line of Eq.~(\ref{eq:anisotropicEM}) and observing that the eccentricity parameter drops out. The resulting Fermi surface represents an ellipsoidal (prolate when $M < 1$ and oblate when $M>1$) conduction band with in-plane ($x\text{-}y$) versus out-of-plane ($z$) anisotropy.
All resistivity scaling curves show similar behavior with lower in-plane mass and higher out-of-plane mass resulting in a lower resistivity for all film thicknesses. The bulk collision time or mean free path predominantly affects the large film thickness behavior which is found to approach the bulk resistivity limit.
The resistivity is dominated by grain boundary backscattering for all degrees of anisotropy, increasingly for lower (higher) in-plane (out-of-plane) effective masses. The ratio of resistivity due to grain boundary versus boundary surface scattering reaches a minimum when the film thickness is of the order of the bulk mean free path, the minimum shifting up for a higher (lower) in-plane (out-of-plane) effective mass.

\setlength{\tabcolsep}{4pt}
\begin{table}[tb]
  \begin{tabular}{ l c c c c c }
    \hline \hline
    & & Isotropic & & & \\
    & Texture & quantity & $M_{x,y}$ & $M_z$ & $\EF$ (eV) \\
    \hline
    Cu & Any & $\tau$ & 1.23 & 1.23 & 4.52 \\
    & & $l_0$ & 1.25 & 1.25 & 3.93 \\
    Ru & None & $\tau$ & 1.69 & 1.69 & 3.71 \\
    & & $l_0$ & 1.72 & 1.72 & 3.37 \\
    & [001] & $\tau$ & 1.65 & 1.54 & 5.00 \\
    & & $l_0$ & 1.91 & 1.62 & 2.54 \\
    \hline \hline
  \end{tabular}
  \caption{
    Least-square fit of the diagonal effective mass tensor and Fermi energy for bulk Cu and Ru following the procedure of section~\ref{subsec:fitting_procedure} with Eqs.~(\ref{eq:fit_iso_tau}-\ref{eq:fit_iso_MFP}) and averaging appropriate for the texture under consideration, as explained in Appendix~\ref{sec:appendix2}.
  }
  \label{table:EM}
\end{table}

\begin{figure}[tb]
  \centering
  \includegraphics[width=0.8\linewidth]{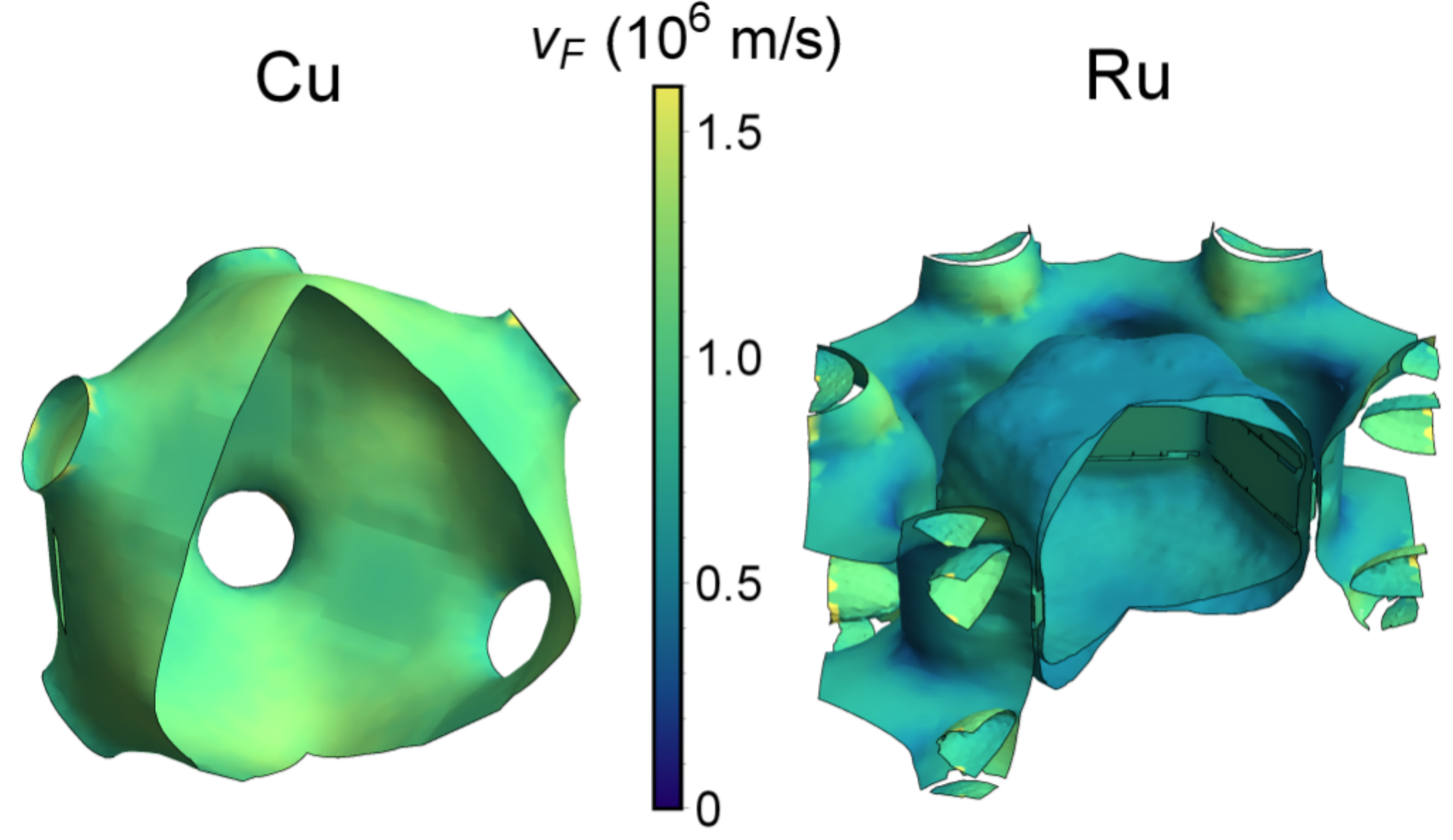}
  \caption{The Fermi surface in the first Brillouin zone is shown for bulk Cu and Ru with the magnitude of the directional Fermi velocity indicated in color.
  }
  \label{fig:bands}
\end{figure}

According to the fitting procedure of section~\ref{subsec:fitting_procedure}, the directional effective masses and Fermi energy were obtained for bulk Cu and Ru, as presented in Table~\ref{table:EM}. For Ru, we consider untextured films for which a spherically averaged Fermi surface with $M_x = M_y = M_z$ is appropriate, as well as [001]-textured films for which in-plane ($x$-$y$) averaging ($M_x = M_y \neq M_z$) of the Fermi surface is appropriate (see Appendix~\ref{sec:appendix2} for details). All the subsequent results depend on the film texture under consideration. For Cu the Fermi surface is quasi-isotropic, hence the fit does not depend on the film texture.
The bulk Fermi surfaces for Cu and Ru (see Fig.~\ref{fig:bands}) were obtained with the computation of the bulk electronic structures using \textit{ab initio} calculations based on the density functional theory implemented in the Quantum-Espresso packages \cite{Giannozzi2009}. Projector augmented wave \cite{Blochl1994} potentials with the Perdew-Burke-Ernzerof generalized gradient \cite{Perdew1996} approximation form of the exchange-correlation functional and a finer Monkhorst-Pack $k$-point sampling grid of 40$\times$40$\times$40 together with kinetic energy cutoff of 80~Ry are used to ensure the total energy to converge up to a tolerance of $10^{-12}$~eV.

\begin{table}[tb]
  \begin{tabular}{ l c c c c c }
    \hline \hline
    & Texture & $\rho_{x\text{-}y}$ ($\mu\Omega\,$cm) & $\rho_z$ ($\mu\Omega \,$cm) & $\tau$ (fs) & $l_0$ (nm) \\
    \hline
    Cu & Any & 1.71 & 1.71 & 38.8 & 43.8 \\
    Ru & None & 7.05 & 7.05 & 9.2 & 6.6 \\
     & [001] & 7.62 & 5.82 & 6.0 & 8.2 \\
    \hline \hline
  \end{tabular}
  \caption{
    Values for the (isotropic) collision time ($\tau$) and mean free path ($l_0$) are listed for Cu and untextured and [001]-textured Ru films, as obtained from the fitted effective masses and Fermi energy in Table~\ref{table:EM},
    the bulk limit of Eqs.~(\ref{eq:cond_isoTau}) and (\ref{eq:cond_isoMFP}), and the experimental resistivity values at 300~K (in-plane: $\rho_{x\text{-}y}$, out-of-plane: $\rho_z$) \cite{bass1982landolt}.
  }
  \label{table:l0_rho0}
\end{table}

Based on the fit for the effective masses and the Fermi energy, the value of $\sigma^\bulk/\tau$ or $\sigma^\bulk/l_0$ can be obtained, assuming an isotropic bulk collision time or an isotropic bulk mean free path, by plugging the obtained values into the bulk limit of Eqs.~(\ref{eq:cond_isoTau}) and (\ref{eq:cond_isoMFP}) respectively. We extract a value for the isotropic collision time or mean free path by fixing the resistivity along the transport direction to the experimental value at room temperature \cite{bass1982landolt}. The obtained values are provided in Table~\ref{table:l0_rho0}.

\begin{figure}[tb]
  \centering
  \subfigure[\ ]{\includegraphics[width=0.8\linewidth]{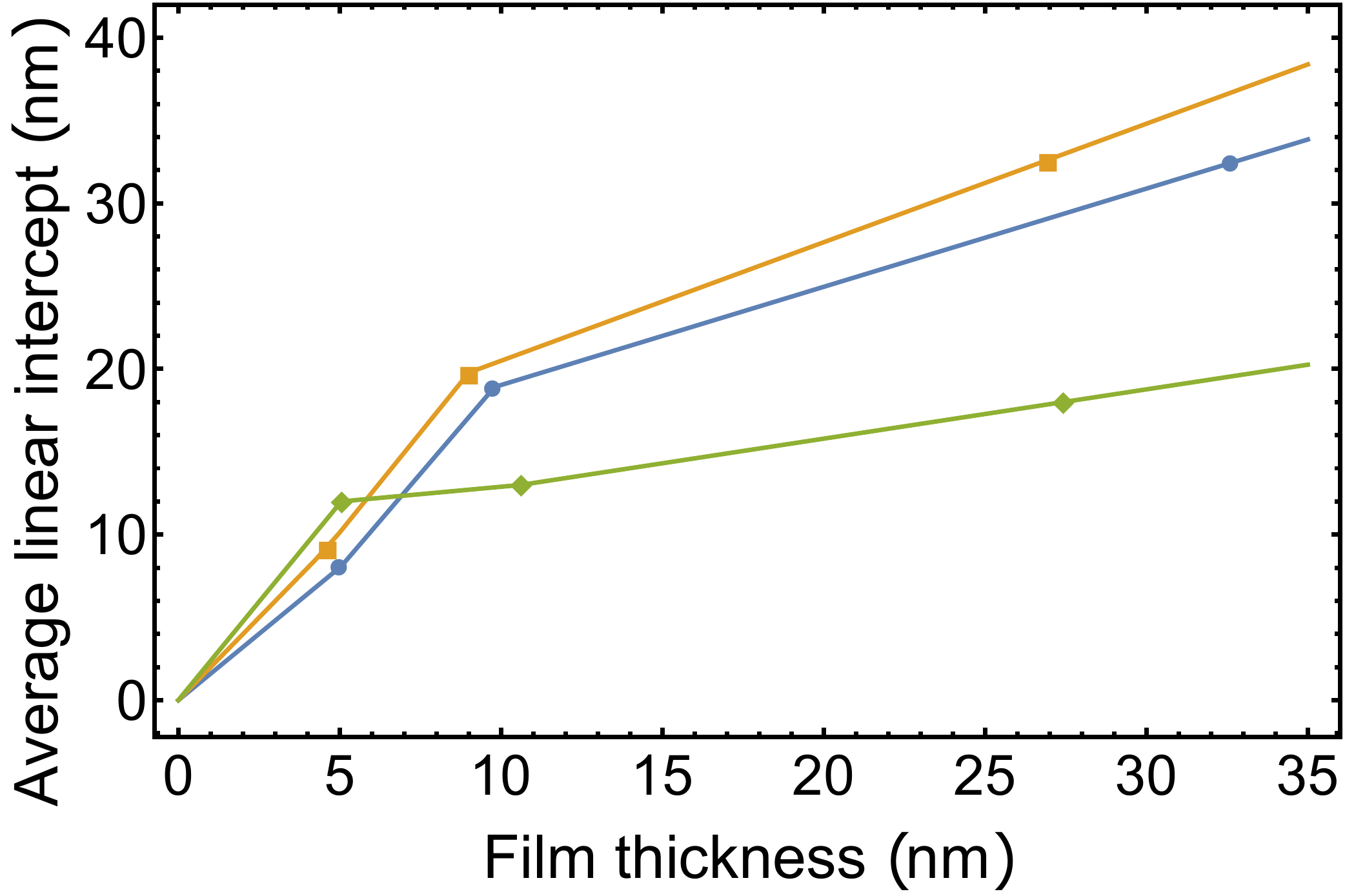}}
  \subfigure[\ ]{\includegraphics[width=0.8\linewidth]{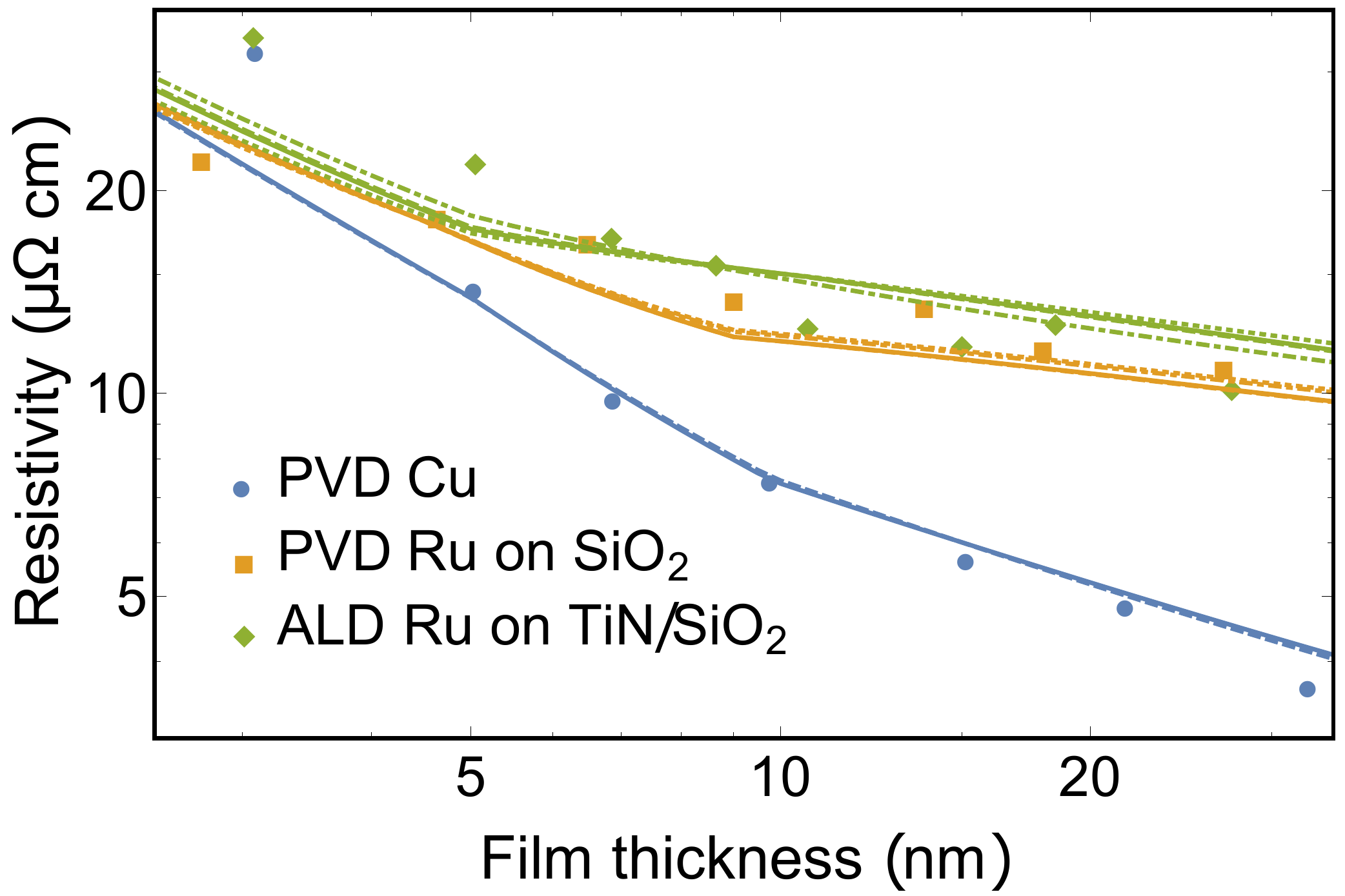}}
  \caption{
    (a) Experimental values for the average linear intercept as a function of the film thickness. Piecewise linear interpolation is considered for the resistivity scaling model.
    (b) Resistivity scaling fits and experimental resistivity data on a log-log scale of PVD Cu, PVD Ru on SiO${}_2$ and ALD Ru on TiN/SiO${}_2$. The fits for $p$ and $R$ are listed in Table~\ref{table:MS_p_R}. They are obtained for Cu and untextured Ru, assuming isotropic $\tau$ (full) or isotropic $l_0$ (dashed), as well as for [001]-textured Ru with isotropic $\tau$ (dotted) or isotropic $l_0$ (dot-dashed).
  }
  \label{fig:grains_res}
\end{figure}

We have now fixed all the parameters that enter the conductivity formulas in Eqs.~(\ref{eq:cond_isoTau}) and (\ref{eq:cond_isoMFP}), based on \textit{ab initio} Fermi surface calculations and calibration with bulk resistivity data, apart from the reflection coefficient $R$, the specularity parameter $p$ and the average linear intercept $d$ between grains.
The average linear intercept has been estimated from TEM images for an appropriate subset of the thin film samples according to the standard method in \cite{standard1996e112}, and the relation $d(t)$ for any thickness $t$ is obtained with linear interpolation, with the addition of a virtual data point ($t=0$, $d=0$), reflecting an imperfect thin film in the limit $t \rightarrow 0$.
Hence, the reflection coefficient $R$ and the specularity parameter $p$ remain as the only two fitting parameters.
We will fit $R$ and $p$ to experimental thickness-dependent resistivity and intercept data of to both Cu and Ru films. Cu films were deposited by physical-vapor deposition (PVD, sputtering) at room temperature on 1.5 nm TaN/SiO2/Si substrates and capped by 1.5 nm of TaN to prevent oxidation of Cu. The films showed an fcc crystalline structure with strong [111] texture \cite{Dutta2017}. PVD Ru films were deposited on SiO${}_2$/Si substrates, leading to strong hexagonal [001] texture \cite{Dutta2017}. Additional Ru films were deposited by atomic layer deposition on TiN/SiO2/Si substrates leading to the absence of any texture in the films, i.e. the films were random hexagonal polycrystals \cite{Popovici2017}. The microstructure of all thin films was found to be near-bamboo-like with only few grain boundaries parallel to the sample surface, such that the grain boundary normals are perpendicular to the out-of-plane-direction, and without any preferred in-plane orientation. Thin film resistivities were obtained from four-point sheet resistance measurements at room temperature as well as the film thickness measured by both x-ray reflectance and Rutherford backscattering spectrometry \cite{Dutta2017}.

\begin{table} [tb]
  \begin{tabular}{ l  c  c  c  c  c }
    \hline \hline
    & & Isotropic & & \\
    & Texture & quantity & $R$ & $p$ & $\sqrt{\textnormal{SSE}}$ ($\mu\Omega\,$cm) \\
    \hline
    PVD Cu & \underline{Any} & \underline{$\tau$} & \underline{0.22} & \underline{0.02} & \underline{0.86} \\
     & & $l_0$ & 0.18 & 0.00 & 0.86 \\
    PVD Ru & None & $\tau$ & 0.45 & 0.99 & 4.72 \\
     & & $l_0$ & 0.43 & 0.94 & 4.71 \\
     & \underline{[001]} & \underline{$\tau$} & \underline{0.48} & \underline{0.96} & \underline{4.08} \\
     & & $l_0$ & 0.37 & 1.00 & 4.25 \\
    ALD Ru & \underline{None} & \underline{$\tau$} & \underline{0.40} & \underline{0.00} & \underline{5.94} \\
     & & $l_0$ & 0.38 & 0.00 & 5.80 \\
     & [001] & $\tau$ & 0.43 & 0.00 & 6.28 \\
     & & $l_0$ & 0.26 & 0.00 & 4.80 \\
    \hline \hline
  \end{tabular}
  \caption{
    Values for the reflection coefficient $R$ and specularity parameter $p$ for the different metal thin films under consideration with the different assumptions for the electronic structure and dominant bulk scattering process. The appropriate set of assumptions for each set of data is underlined.
  }
  \label{table:MS_p_R}
\end{table}

All the fitted parameters for the different sets of data and different assumptions for film texture and bulk scattering (isotropic collision time or isotropic mean free path) are listed in Table~\ref{table:MS_p_R} and the resulting resistivity curves are shown in Fig.~\ref{fig:grains_res}~(b). The steep resistivity increase for PVD Cu and PVD Ru below a film thickness of 7~nm could not be fitted satisfactorily with any set of parameters $R$ and $p$. Therefore, the presented fits for the PVD data have been obtained without considering the data point for thickness around 3~nm (still undershooting the PVD Ru data point around 5~nm thickness systematically). Even though different film textures and bulk scattering properties underlie the different resistivity scaling curves of a given data set, the curves are in very close agreement.

\begin{figure}[tb]
  \centering
  \includegraphics[width=0.85\columnwidth]{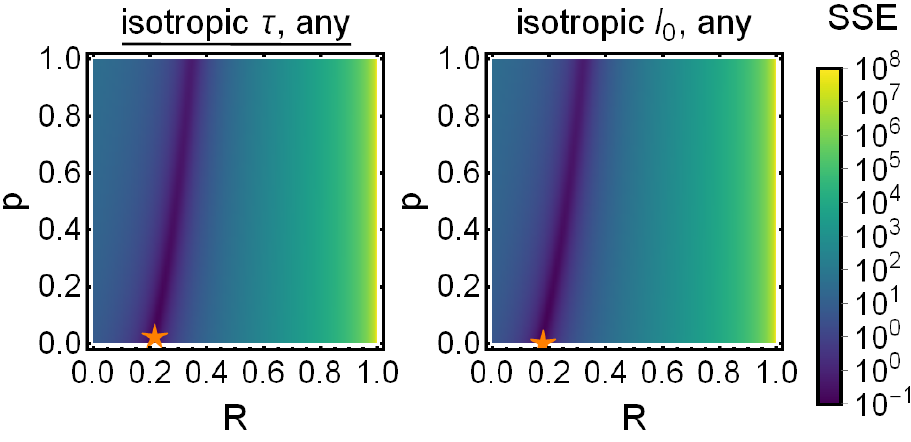}
  \caption{
    The SSE of the fit of (a) Eq.~(\ref{eq:cond_isoTau}) (b) Eq.~(\ref{eq:cond_isoMFP}) to the PVD Cu data of Fig.~\ref{fig:grains_res} is shown for $0 \leq R$, $p \leq 1$. The minimum is indicated with a star.
  }
  \label{fig:SSE_Cu}
\end{figure}

\begin{figure}[tb]
  \centering
  \subfigure[\ ]{\includegraphics[width=0.85\columnwidth]{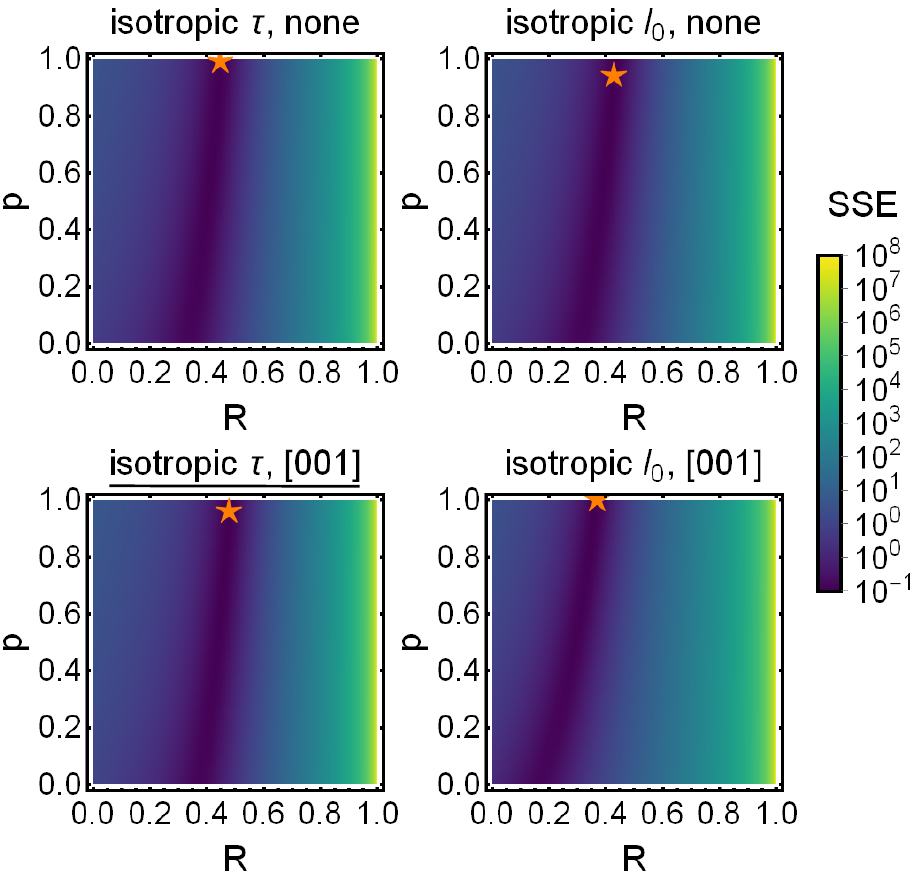}}
  \hspace{0.1\columnwidth}
  \subfigure[\ ]{\includegraphics[width=0.85\columnwidth]{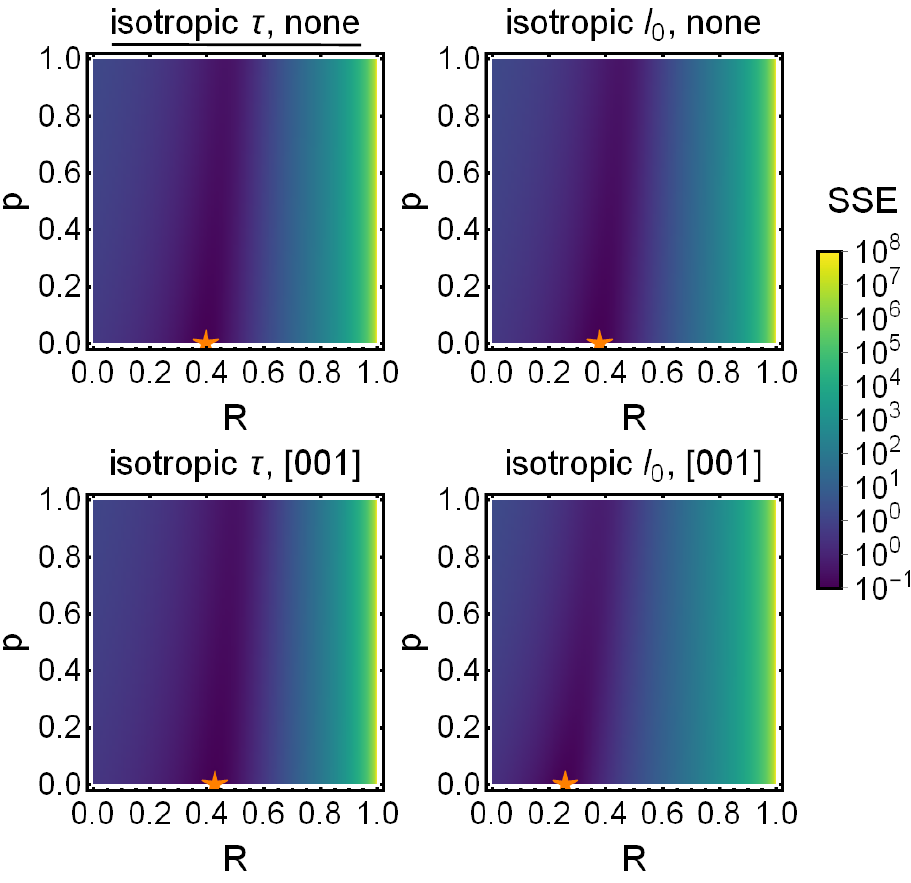}}
  \caption{
    The SSE of the fit of Eq.~(\ref{eq:cond_isoTau}) for isotropic collision time (left column) and Eq.~(\ref{eq:cond_isoMFP}) for isotropic mean free path (right column) to the (a) PVD Ru on SiO${}_2$ (b) ALD Ru on TiN/SiO${}_2$ data of Fig.~\ref{fig:grains_res} is shown for $0 \leq R$, $p \leq 1$, assuming an untextured (top row) and a [001]-textured film (bottom row). The minimum is indicated with a star.
  }
  \label{fig:SSE_Ru}
\end{figure}

The SSE of the full range of $R$ and $p$ is shown for all the combinations of film texture and bulk scattering assumptions for the different fits in Figs.~\ref{fig:SSE_Cu}-\ref{fig:SSE_Ru}, with the results of Table~\ref{table:MS_p_R} indicated. The fit is very robust for the reflection coefficient $R$, unlike for the specularity parameter $p$. Furthermore, slight differences between the reflection coefficients can be identified when considering a [001]-textured thin film with isotropic collision time or mean free path for PVD and ALD Ru. Assuming an isotropic collision time (mean free path), increases (decreases) the obtained reflection coefficient with respect to the untextured film assumption, for which the assumption regarding the bulk scattering isotropy is of little importance. The specularity parameter is in general barely affected.

\section{Discussion} \label{sec:discussion}
It is evident from the results that conduction band anisotropy can have an impact on thin film resistivity scaling. The resistivity curves with different degrees of in-plane versus out-of-plane anisotropy in Fig.~\ref{fig:res_scaling} clearly show the impact on resistivity scaling, the lowest resistivity being realized by the lowest in-plane and highest out-of-plane effective masses. As such, the electrons suffer less from boundary scattering and are more resilient to grain boundary scattering.
The impact is more pronounced in case of an isotropic collision time as compared to an isotropic mean free path and this can be understood by having a closer look at the conductivity expressions in Eqs.~(\ref{eq:cond_GB_BS}), (\ref{eq:cond_isoTau}) and (\ref{eq:cond_isoMFP}).
In case of an isotropic collision time, both the effective mass along the transport direction and the parameter for grain boundary scattering are renormalized w.r.t. the Mayadas-Shatzkes formula ($m_\e \rightarrow m_x$, $\alpha \rightarrow \beta$). As the latter is not renormalized in case of an isotropic mean free path, the impact of anisotropy is more limited.

When the average linear intercept is comparable to the film thickness, grain boundary scattering is typically more detrimental to the conductivity than boundary surface scattering. This feature can clearly be seen in Fig.~\ref{fig:res_scaling} and has also been observed before \cite{Sun2009, Dutta2017}.
This behavior also shows from Figs.~\ref{fig:SSE_Cu} and \ref{fig:SSE_Ru}, with all fits showing a strong correlation along $R(p) = R(0) + Bp$ with $B \gg 1$, resulting in a quite robust reflection coefficient $R$ and a specularity parameter $p$ which can easily swing up or down for minor variations in the data or for different assumptions regarding the film texture and bulk scattering properties.
The physical reason is that backscattering at grain boundaries is maximally detrimental to the electron transport velocity, which gets completely reversed by the scattering event ($v_x \rightarrow -v_x$). Conversely, a diffusive boundary scattering event reorients the velocity arbitrarily, possibly retaining or even enhancing the transport velocity of a conduction electron. One should therefore be careful drawing any conclusion from the optimal value of $p$ in these fits (see Table~\ref{table:MS_p_R}), as it carries little physical meaning.

While the quality of the fit to thickness-dependent resistivity data and the value of the specularity parameter barely depend on the different assumptions for film texture and bulk scattering (isotropy of bulk collision time or mean free path), the obtained value of the grain boundary reflection coefficient can be affected. When attempting to draw meaningful conclusions on grain boundary scattering from the precise value of the reflection coefficient, one should also consider the film texture and degree of conduction band anisotropy as well as the characteristics of the dominant bulk scattering mechanism before proceeding with the appropriate fitting procedure. These complications arise only when one considers metals with a high degree of anisotropy, such that one has to go beyond the standard Mayadas-Shatzkes formula. For Cu, the conduction band is nearly isotropic and the different assumptions lead to results that are in close agreement.

All the resistivity measurements in the fits were performed at room temperature, so we expect the (acoustic) phonons to provide the dominant scattering mechanism in the bulk regime, rendering the isotropic collision time assumption and the corresponding $R$ parameters meaningful. Furthermore, the [001]-textured film assumption should apply to PVD Ru while the results with a untextured film assumption should be considered for ALD Ru, even though this does not provide the best overall fit. The appropriateness of the fit is not reflected in the quality of the fit, but it certainly dismisses the validity of certain results under different assumptions. For example, the reflection coefficient would turn out to be significantly larger for the Ru samples if the [001]-texture with isotropic bulk mean free path would be the appropriate assumption.

A returning problem of the Mayadas-Shatzkes model and its extension for anisotropic conduction bands is the underestimation of the resistivity for small film thicknesses (for PVD Cu and PVD Ru on SiO${}_2$ in our case) \cite{Zheng2017}. While it is hard to guarantee film continuity for the samples at very low thicknesses, several limitations and approximations underlying the Mayadas-Shatzkes model and its extension (including the consideration of a three-dimensional wave vector, the phenomenological treatment of boundary surface scattering and the perturbative treatment of grain boundary scattering) could also be (partially) responsible for a systematic underestimation (already remarked by Choi \textit{et al.} \cite{choi2014failure}). A recent validity analysis showed that this is certainly the case for a perturbative treatment of boundary surface scattering due to surface roughness and grain boundary scattering for extremely narrow nanowires \cite{Moors2016}.
Additionally, the consideration of a single (diagonal) effective mass (tensor) could be too approximate for complicated Fermi surfaces such as for Ru. A full numerical simulation of Eq.~(\ref{eq:BTE_Surface}) with the correct Fermi surface, originating from multiple bands, should be able to rule this out. A recent study by Zheng \textit{et al.} for W thin films, while limited to thin films with fully diffusive boundary surface scattering and without grain boundary scattering, goes in this direction \cite{Zheng2017B}. Furthermore, one could refine the simplistic perpendicular delta-function barriers for grain boundary scattering. Atomistic studies for extremely small nanowires with few grain boundaries with different orientations have been done \cite{Cesar2014, Cesar2016, Lanzillo2017}, but require a full numerical treatment for which metal thin films with realistic grain profiles are too complicated.
In recent work, Li \textit{et al.} introduced a phenomenological correction for conduction band anisotropy and argued that it was responsible for the sharper resistivity increase at small thicknesses for Os films \cite{Li2017}. Our results do not agree with this as they cannot explain an underestimation of the resistivity for small thicknesses merely based on conduction band anisotropy. Another issue that might explain the deviations is the lack of data on average linear intercepts below 5~nm thickness. Since grain boundary scattering is dominant, the resistivity curve will be strongly dependent on the value of $d$, hence an accurately determined value of the average linear intercept is crucial for the fitting procedure. The linear interpolation that we employed might be too limited.

\section{Conclusion} \label{sec:conclusion}
The semiclassical resistivity scaling formula derived by Mayadas and Shatzkes for metal thin films with grain boundary and boundary surface scattering is extended to account for conduction band anisotropy. Apart from a dependency on the film texture, this extension introduces an additional dependency on the anisotropy of the dominant bulk scattering mechanism. An explicit formulation for the thickness-dependent resistivity is derived and presented for two limit cases: bulk scattering with an isotropic collision time and with an isotropic mean free path.

A systematic procedure is presented to fit a (or multiple) highly anisotropic conduction band(s) of a metal of choice to an ellipsoidal Fermi surface and then demonstrated for Cu and Ru, respectively nearly isotropic and highly anisotropic. This procedure allows us to systematically fit the grain boundary reflection coefficient and specularity parameter of the film boundary surfaces for textured and untextured metal thin films with anisotropic conduction bands. A dependency of the reflection coefficient on the film texture and the bulk scattering characteristics is observed, while the obtained specularity parameter is barely unaffected and moreover physically insignificant due to the resistivity contribution of boundary surface scattering being much weaker. While conduction band anisotropy has been suggested before as a possible explanation for deviations from Mayadas-Shatzkes (e.g. a steeper resistivity increase for very small film thicknesses), this hypothesis does not follow from our model. The main consequence of considering conduction band anisotropy is a renormalization of the reflection coefficient, without significantly adjusting the quality of the fit to experimental thickness-dependent resistivity data.

\begin{acknowledgments}
The authors acknowledge the support by the National Research Fund Luxembourg (ATTRACT Grant No.~7556175).
\end{acknowledgments}

\bibliography{2017_DeClercq_Thin_Film_arXiv}{}

\begin{thebibliography}{40}%
\makeatletter
\providecommand \@ifxundefined [1]{%
 \@ifx{#1\undefined}
}%
\providecommand \@ifnum [1]{%
 \ifnum #1\expandafter \@firstoftwo
 \else \expandafter \@secondoftwo
 \fi
}%
\providecommand \@ifx [1]{%
 \ifx #1\expandafter \@firstoftwo
 \else \expandafter \@secondoftwo
 \fi
}%
\providecommand \natexlab [1]{#1}%
\providecommand \enquote  [1]{``#1''}%
\providecommand \bibnamefont  [1]{#1}%
\providecommand \bibfnamefont [1]{#1}%
\providecommand \citenamefont [1]{#1}%
\providecommand \href@noop [0]{\@secondoftwo}%
\providecommand \href [0]{\begingroup \@sanitize@url \@href}%
\providecommand \@href[1]{\@@startlink{#1}\@@href}%
\providecommand \@@href[1]{\endgroup#1\@@endlink}%
\providecommand \@sanitize@url [0]{\catcode `\\12\catcode `\$12\catcode
  `\&12\catcode `\#12\catcode `\^12\catcode `\_12\catcode `\%12\relax}%
\providecommand \@@startlink[1]{}%
\providecommand \@@endlink[0]{}%
\providecommand \url  [0]{\begingroup\@sanitize@url \@url }%
\providecommand \@url [1]{\endgroup\@href {#1}{\urlprefix }}%
\providecommand \urlprefix  [0]{URL }%
\providecommand \Eprint [0]{\href }%
\providecommand \doibase [0]{http://dx.doi.org/}%
\providecommand \selectlanguage [0]{\@gobble}%
\providecommand \bibinfo  [0]{\@secondoftwo}%
\providecommand \bibfield  [0]{\@secondoftwo}%
\providecommand \translation [1]{[#1]}%
\providecommand \BibitemOpen [0]{}%
\providecommand \bibitemStop [0]{}%
\providecommand \bibitemNoStop [0]{.\EOS\space}%
\providecommand \EOS [0]{\spacefactor3000\relax}%
\providecommand \BibitemShut  [1]{\csname bibitem#1\endcsname}%
\let\auto@bib@innerbib\@empty
%</preamble>
\bibitem [{\citenamefont {Fuchs}(1938)}]{fuchs1938conductivity}%
  \BibitemOpen
  \bibfield  {author} {\bibinfo {author} {\bibfnamefont {K.}~\bibnamefont
  {Fuchs}},\ }\href@noop {} {\bibfield  {journal} {\bibinfo  {journal}
  {Proceedings of Cambridge Philosophical Society}\ }\textbf {\bibinfo {volume}
  {34}},\ \bibinfo {pages} {100} (\bibinfo {year} {1938})}\BibitemShut
  {NoStop}%
\bibitem [{\citenamefont {Sondheimer}(1952)}]{sondheimer1952mean}%
  \BibitemOpen
  \bibfield  {author} {\bibinfo {author} {\bibfnamefont {E.~H.}\ \bibnamefont
  {Sondheimer}},\ }\href@noop {} {\bibfield  {journal} {\bibinfo  {journal}
  {Advances in Physics}\ }\textbf {\bibinfo {volume} {1}},\ \bibinfo {pages}
  {1} (\bibinfo {year} {1952})}\BibitemShut {NoStop}%
\bibitem [{\citenamefont {Mayadas}\ and\ \citenamefont
  {Shatzkes}(1970)}]{mayadas1970electrical}%
  \BibitemOpen
  \bibfield  {author} {\bibinfo {author} {\bibfnamefont {A.}~\bibnamefont
  {Mayadas}}\ and\ \bibinfo {author} {\bibfnamefont {M.}~\bibnamefont
  {Shatzkes}},\ }\href@noop {} {\bibfield  {journal} {\bibinfo  {journal}
  {Physical Review B}\ }\textbf {\bibinfo {volume} {1}},\ \bibinfo {pages}
  {1382} (\bibinfo {year} {1970})}\BibitemShut {NoStop}%
\bibitem [{\citenamefont {Soffer}(1967)}]{soffer1967statistical}%
  \BibitemOpen
  \bibfield  {author} {\bibinfo {author} {\bibfnamefont {S.~B.}\ \bibnamefont
  {Soffer}},\ }\href@noop {} {\bibfield  {journal} {\bibinfo  {journal}
  {Journal of Applied Physics}\ }\textbf {\bibinfo {volume} {38}},\ \bibinfo
  {pages} {1710} (\bibinfo {year} {1967})}\BibitemShut {NoStop}%
\bibitem [{\citenamefont {Trivedi}\ and\ \citenamefont
  {Ashcroft}(1988)}]{trivedi1988quantum}%
  \BibitemOpen
  \bibfield  {author} {\bibinfo {author} {\bibfnamefont {N.}~\bibnamefont
  {Trivedi}}\ and\ \bibinfo {author} {\bibfnamefont {N.~W.}\ \bibnamefont
  {Ashcroft}},\ }\href@noop {} {\bibfield  {journal} {\bibinfo  {journal}
  {Physical Review B}\ }\textbf {\bibinfo {volume} {38}},\ \bibinfo {pages}
  {12298} (\bibinfo {year} {1988})}\BibitemShut {NoStop}%
\bibitem [{\citenamefont {Fishman}\ and\ \citenamefont
  {Calecki}(1989)}]{fishman1989surface}%
  \BibitemOpen
  \bibfield  {author} {\bibinfo {author} {\bibfnamefont {G.}~\bibnamefont
  {Fishman}}\ and\ \bibinfo {author} {\bibfnamefont {D.}~\bibnamefont
  {Calecki}},\ }\href@noop {} {\bibfield  {journal} {\bibinfo  {journal}
  {Physical Review Letters}\ }\textbf {\bibinfo {volume} {62}},\ \bibinfo
  {pages} {1302} (\bibinfo {year} {1989})}\BibitemShut {NoStop}%
\bibitem [{\citenamefont {Zhang}\ and\ \citenamefont
  {Butler}(1995)}]{zhang1995conductivity}%
  \BibitemOpen
  \bibfield  {author} {\bibinfo {author} {\bibfnamefont {X.-G.}\ \bibnamefont
  {Zhang}}\ and\ \bibinfo {author} {\bibfnamefont {W.~H.}\ \bibnamefont
  {Butler}},\ }\href@noop {} {\bibfield  {journal} {\bibinfo  {journal}
  {Physical Review B}\ }\textbf {\bibinfo {volume} {51}},\ \bibinfo {pages}
  {10085} (\bibinfo {year} {1995})}\BibitemShut {NoStop}%
\bibitem [{\citenamefont {Palasantzas}(1998)}]{palasantzas1998surface}%
  \BibitemOpen
  \bibfield  {author} {\bibinfo {author} {\bibfnamefont {G.}~\bibnamefont
  {Palasantzas}},\ }\href@noop {} {\bibfield  {journal} {\bibinfo  {journal}
  {Physical Review B}\ }\textbf {\bibinfo {volume} {58}},\ \bibinfo {pages}
  {9685} (\bibinfo {year} {1998})}\BibitemShut {NoStop}%
\bibitem [{\citenamefont {Meyerovich}\ and\ \citenamefont
  {Ponomarev}(2002)}]{meyerovich2002surface}%
  \BibitemOpen
  \bibfield  {author} {\bibinfo {author} {\bibfnamefont {A.~E.}\ \bibnamefont
  {Meyerovich}}\ and\ \bibinfo {author} {\bibfnamefont {I.~V.}\ \bibnamefont
  {Ponomarev}},\ }\href@noop {} {\bibfield  {journal} {\bibinfo  {journal}
  {Physical Review B}\ }\textbf {\bibinfo {volume} {65}},\ \bibinfo {pages}
  {155413} (\bibinfo {year} {2002})}\BibitemShut {NoStop}%
\bibitem [{\citenamefont {Rickman}\ and\ \citenamefont
  {Barmak}(2012)}]{Rickman2012}%
  \BibitemOpen
  \bibfield  {author} {\bibinfo {author} {\bibfnamefont {J.~M.}\ \bibnamefont
  {Rickman}}\ and\ \bibinfo {author} {\bibfnamefont {K.}~\bibnamefont
  {Barmak}},\ }\href@noop {} {\bibfield  {journal} {\bibinfo  {journal}
  {Journal of Applied Physics}\ }\textbf {\bibinfo {volume} {112}},\ \bibinfo
  {pages} {013704} (\bibinfo {year} {2012})}\BibitemShut {NoStop}%
\bibitem [{\citenamefont {Moors}\ \emph {et~al.}(2016)\citenamefont {Moors},
  \citenamefont {Sor{\'{e}}e},\ and\ \citenamefont {Magnus}}]{Moors2016}%
  \BibitemOpen
  \bibfield  {author} {\bibinfo {author} {\bibfnamefont {K.}~\bibnamefont
  {Moors}}, \bibinfo {author} {\bibfnamefont {B.}~\bibnamefont {Sor{\'{e}}e}},
  \ and\ \bibinfo {author} {\bibfnamefont {W.}~\bibnamefont {Magnus}},\
  }\href@noop {} {\bibfield  {journal} {\bibinfo  {journal} {Journal of Physics
  Condensed Matter}\ }\textbf {\bibinfo {volume} {28}},\ \bibinfo {pages}
  {365302} (\bibinfo {year} {2016})}\BibitemShut {NoStop}%
\bibitem [{\citenamefont {Munoz}\ and\ \citenamefont
  {Arenas}(2017)}]{Munoz2017}%
  \BibitemOpen
  \bibfield  {author} {\bibinfo {author} {\bibfnamefont {R.~C.}\ \bibnamefont
  {Munoz}}\ and\ \bibinfo {author} {\bibfnamefont {C.}~\bibnamefont {Arenas}},\
  }\href@noop {} {\bibfield  {journal} {\bibinfo  {journal} {Applied Physics
  Reviews}\ }\textbf {\bibinfo {volume} {4}},\ \bibinfo {pages} {011102}
  (\bibinfo {year} {2017})}\BibitemShut {NoStop}%
\bibitem [{\citenamefont {De~Vries}(1988)}]{de1988temperature}%
  \BibitemOpen
  \bibfield  {author} {\bibinfo {author} {\bibfnamefont {J.~W.~C.}\
  \bibnamefont {De~Vries}},\ }\href@noop {} {\bibfield  {journal} {\bibinfo
  {journal} {Thin Solid Films}\ }\textbf {\bibinfo {volume} {167}},\ \bibinfo
  {pages} {25} (\bibinfo {year} {1988})}\BibitemShut {NoStop}%
\bibitem [{\citenamefont {Liu}\ \emph {et~al.}(2001)\citenamefont {Liu},
  \citenamefont {Zhao}, \citenamefont {Ramanath}, \citenamefont {Murarka},\
  and\ \citenamefont {Wang}}]{liu2001thickness}%
  \BibitemOpen
  \bibfield  {author} {\bibinfo {author} {\bibfnamefont {H.-D.}\ \bibnamefont
  {Liu}}, \bibinfo {author} {\bibfnamefont {Y.-P.}\ \bibnamefont {Zhao}},
  \bibinfo {author} {\bibfnamefont {G.}~\bibnamefont {Ramanath}}, \bibinfo
  {author} {\bibfnamefont {S.~P.}\ \bibnamefont {Murarka}}, \ and\ \bibinfo
  {author} {\bibfnamefont {G.-C.}\ \bibnamefont {Wang}},\ }\href@noop {}
  {\bibfield  {journal} {\bibinfo  {journal} {Thin Solid Films}\ }\textbf
  {\bibinfo {volume} {384}},\ \bibinfo {pages} {151} (\bibinfo {year}
  {2001})}\BibitemShut {NoStop}%
\bibitem [{\citenamefont {Tay}\ \emph {et~al.}(2005)\citenamefont {Tay},
  \citenamefont {Li},\ and\ \citenamefont {Wu}}]{tay2005electrical}%
  \BibitemOpen
  \bibfield  {author} {\bibinfo {author} {\bibfnamefont {M.}~\bibnamefont
  {Tay}}, \bibinfo {author} {\bibfnamefont {K.}~\bibnamefont {Li}}, \ and\
  \bibinfo {author} {\bibfnamefont {Y.}~\bibnamefont {Wu}},\ }\href@noop {}
  {\bibfield  {journal} {\bibinfo  {journal} {Journal of Vacuum Science \&
  Technology B}\ }\textbf {\bibinfo {volume} {23}},\ \bibinfo {pages} {1412}
  (\bibinfo {year} {2005})}\BibitemShut {NoStop}%
\bibitem [{\citenamefont {Camacho}\ and\ \citenamefont
  {Oliva}(2006)}]{camacho2006surface}%
  \BibitemOpen
  \bibfield  {author} {\bibinfo {author} {\bibfnamefont {J.~M.}\ \bibnamefont
  {Camacho}}\ and\ \bibinfo {author} {\bibfnamefont {A.}~\bibnamefont
  {Oliva}},\ }\href@noop {} {\bibfield  {journal} {\bibinfo  {journal} {Thin
  Solid Films}\ }\textbf {\bibinfo {volume} {515}},\ \bibinfo {pages} {1881}
  (\bibinfo {year} {2006})}\BibitemShut {NoStop}%
\bibitem [{\citenamefont {Sun}\ \emph {et~al.}(2010)\citenamefont {Sun},
  \citenamefont {Yao}, \citenamefont {Warren}, \citenamefont {Barmak},
  \citenamefont {Toney}, \citenamefont {Peale},\ and\ \citenamefont
  {Coffey}}]{sun2010surface}%
  \BibitemOpen
  \bibfield  {author} {\bibinfo {author} {\bibfnamefont {T.}~\bibnamefont
  {Sun}}, \bibinfo {author} {\bibfnamefont {B.}~\bibnamefont {Yao}}, \bibinfo
  {author} {\bibfnamefont {A.~P.}\ \bibnamefont {Warren}}, \bibinfo {author}
  {\bibfnamefont {K.}~\bibnamefont {Barmak}}, \bibinfo {author} {\bibfnamefont
  {M.~F.}\ \bibnamefont {Toney}}, \bibinfo {author} {\bibfnamefont {R.~E.}\
  \bibnamefont {Peale}}, \ and\ \bibinfo {author} {\bibfnamefont {K.~R.}\
  \bibnamefont {Coffey}},\ }\href@noop {} {\bibfield  {journal} {\bibinfo
  {journal} {Physical Review B}\ }\textbf {\bibinfo {volume} {81}},\ \bibinfo
  {pages} {155454} (\bibinfo {year} {2010})}\BibitemShut {NoStop}%
\bibitem [{\citenamefont {Chawla}\ \emph {et~al.}(2011)\citenamefont {Chawla},
  \citenamefont {Gstrein}, \citenamefont {O’Brien}, \citenamefont {Clarke},\
  and\ \citenamefont {Gall}}]{chawla2011electron}%
  \BibitemOpen
  \bibfield  {author} {\bibinfo {author} {\bibfnamefont {J.~S.}\ \bibnamefont
  {Chawla}}, \bibinfo {author} {\bibfnamefont {F.}~\bibnamefont {Gstrein}},
  \bibinfo {author} {\bibfnamefont {K.~P.}\ \bibnamefont {O’Brien}}, \bibinfo
  {author} {\bibfnamefont {J.~S.}\ \bibnamefont {Clarke}}, \ and\ \bibinfo
  {author} {\bibfnamefont {D.}~\bibnamefont {Gall}},\ }\href@noop {} {\bibfield
   {journal} {\bibinfo  {journal} {Physical Review B}\ }\textbf {\bibinfo
  {volume} {84}},\ \bibinfo {pages} {235423} (\bibinfo {year}
  {2011})}\BibitemShut {NoStop}%
\bibitem [{\citenamefont {Sankaran}\ \emph {et~al.}(2015)\citenamefont
  {Sankaran}, \citenamefont {Clima}, \citenamefont {Mees},\ and\ \citenamefont
  {Pourtois}}]{sankaran2015exploring}%
  \BibitemOpen
  \bibfield  {author} {\bibinfo {author} {\bibfnamefont {K.}~\bibnamefont
  {Sankaran}}, \bibinfo {author} {\bibfnamefont {S.}~\bibnamefont {Clima}},
  \bibinfo {author} {\bibfnamefont {M.}~\bibnamefont {Mees}}, \ and\ \bibinfo
  {author} {\bibfnamefont {G.}~\bibnamefont {Pourtois}},\ }\href@noop {}
  {\bibfield  {journal} {\bibinfo  {journal} {ECS Journal of Solid State
  Science and Technology}\ }\textbf {\bibinfo {volume} {4}},\ \bibinfo {pages}
  {N3127} (\bibinfo {year} {2015})}\BibitemShut {NoStop}%
\bibitem [{\citenamefont {Gall}(2016)}]{gall2016electron}%
  \BibitemOpen
  \bibfield  {author} {\bibinfo {author} {\bibfnamefont {D.}~\bibnamefont
  {Gall}},\ }\href@noop {} {\bibfield  {journal} {\bibinfo  {journal} {Journal
  of Applied Physics}\ }\textbf {\bibinfo {volume} {119}},\ \bibinfo {pages}
  {085101} (\bibinfo {year} {2016})}\BibitemShut {NoStop}%
\bibitem [{\citenamefont {Dutta}\ \emph {et~al.}(2017)\citenamefont {Dutta},
  \citenamefont {Sankaran}, \citenamefont {Moors}, \citenamefont {Pourtois},
  \citenamefont {{Van Elshocht}}, \citenamefont {B{\"{o}}mmels}, \citenamefont
  {Vandervorst}, \citenamefont {T\H{o}kei},\ and\ \citenamefont
  {Adelmann}}]{Dutta2017}%
  \BibitemOpen
  \bibfield  {author} {\bibinfo {author} {\bibfnamefont {S.}~\bibnamefont
  {Dutta}}, \bibinfo {author} {\bibfnamefont {K.}~\bibnamefont {Sankaran}},
  \bibinfo {author} {\bibfnamefont {K.}~\bibnamefont {Moors}}, \bibinfo
  {author} {\bibfnamefont {G.}~\bibnamefont {Pourtois}}, \bibinfo {author}
  {\bibfnamefont {S.}~\bibnamefont {{Van Elshocht}}}, \bibinfo {author}
  {\bibfnamefont {J.}~\bibnamefont {B{\"{o}}mmels}}, \bibinfo {author}
  {\bibfnamefont {W.}~\bibnamefont {Vandervorst}}, \bibinfo {author}
  {\bibfnamefont {Z.}~\bibnamefont {T\H{o}kei}}, \ and\ \bibinfo {author}
  {\bibfnamefont {C.}~\bibnamefont {Adelmann}},\ }\href@noop {} {\bibfield
  {journal} {\bibinfo  {journal} {Journal of Applied Physics}\ }\textbf
  {\bibinfo {volume} {122}},\ \bibinfo {pages} {025107} (\bibinfo {year}
  {2017})}\BibitemShut {NoStop}%
\bibitem [{\citenamefont {Choi}\ \emph {et~al.}(2014)\citenamefont {Choi},
  \citenamefont {Liu}, \citenamefont {Schelling}, \citenamefont {Coffey},\ and\
  \citenamefont {Barmak}}]{choi2014failure}%
  \BibitemOpen
  \bibfield  {author} {\bibinfo {author} {\bibfnamefont {D.}~\bibnamefont
  {Choi}}, \bibinfo {author} {\bibfnamefont {X.}~\bibnamefont {Liu}}, \bibinfo
  {author} {\bibfnamefont {P.~K.}\ \bibnamefont {Schelling}}, \bibinfo {author}
  {\bibfnamefont {K.~R.}\ \bibnamefont {Coffey}}, \ and\ \bibinfo {author}
  {\bibfnamefont {K.}~\bibnamefont {Barmak}},\ }\href@noop {} {\bibfield
  {journal} {\bibinfo  {journal} {Journal of Applied Physics}\ }\textbf
  {\bibinfo {volume} {115}},\ \bibinfo {pages} {104308} (\bibinfo {year}
  {2014})}\BibitemShut {NoStop}%
\bibitem [{\citenamefont {Jones}\ \emph {et~al.}(2015)\citenamefont {Jones},
  \citenamefont {Sanchez-Soares}, \citenamefont {Plombon}, \citenamefont
  {Kaushik}, \citenamefont {Nagle}, \citenamefont {Clarke},\ and\ \citenamefont
  {Greer}}]{Jones2015}%
  \BibitemOpen
  \bibfield  {author} {\bibinfo {author} {\bibfnamefont {S.~L.~T.}\
  \bibnamefont {Jones}}, \bibinfo {author} {\bibfnamefont {A.}~\bibnamefont
  {Sanchez-Soares}}, \bibinfo {author} {\bibfnamefont {J.~J.}\ \bibnamefont
  {Plombon}}, \bibinfo {author} {\bibfnamefont {A.~P.}\ \bibnamefont
  {Kaushik}}, \bibinfo {author} {\bibfnamefont {R.~E.}\ \bibnamefont {Nagle}},
  \bibinfo {author} {\bibfnamefont {J.~S.}\ \bibnamefont {Clarke}}, \ and\
  \bibinfo {author} {\bibfnamefont {J.~C.}\ \bibnamefont {Greer}},\ }\href@noop
  {} {\bibfield  {journal} {\bibinfo  {journal} {Physical Review B}\ }\textbf
  {\bibinfo {volume} {92}},\ \bibinfo {pages} {115413} (\bibinfo {year}
  {2015})}\BibitemShut {NoStop}%
\bibitem [{\citenamefont {Hegde}\ \emph {et~al.}(2016)\citenamefont {Hegde},
  \citenamefont {Bowen},\ and\ \citenamefont {Rodder}}]{Hegde2016}%
  \BibitemOpen
  \bibfield  {author} {\bibinfo {author} {\bibfnamefont {G.}~\bibnamefont
  {Hegde}}, \bibinfo {author} {\bibfnamefont {R.~C.}\ \bibnamefont {Bowen}}, \
  and\ \bibinfo {author} {\bibfnamefont {M.~S.}\ \bibnamefont {Rodder}},\
  }\href@noop {} {\bibfield  {journal} {\bibinfo  {journal} {Applied Physics
  Letters}\ }\textbf {\bibinfo {volume} {109}},\ \bibinfo {pages} {193106}
  (\bibinfo {year} {2016})}\BibitemShut {NoStop}%
\bibitem [{\citenamefont {Lanzillo}(2017)}]{Lanzillo2017}%
  \BibitemOpen
  \bibfield  {author} {\bibinfo {author} {\bibfnamefont {N.~A.}\ \bibnamefont
  {Lanzillo}},\ }\href@noop {} {\bibfield  {journal} {\bibinfo  {journal}
  {Journal of Applied Physics}\ }\textbf {\bibinfo {volume} {121}},\ \bibinfo
  {pages} {175104} (\bibinfo {year} {2017})}\BibitemShut {NoStop}%
\bibitem [{\citenamefont {Li}\ \emph {et~al.}(2017)\citenamefont {Li},
  \citenamefont {Zhang}, \citenamefont {Ma}, \citenamefont {Zhang},
  \citenamefont {Yi},\ and\ \citenamefont {Pan}}]{Li2017}%
  \BibitemOpen
  \bibfield  {author} {\bibinfo {author} {\bibfnamefont {S.~L.}\ \bibnamefont
  {Li}}, \bibinfo {author} {\bibfnamefont {Q.~Y.}\ \bibnamefont {Zhang}},
  \bibinfo {author} {\bibfnamefont {C.~Y.}\ \bibnamefont {Ma}}, \bibinfo
  {author} {\bibfnamefont {C.}~\bibnamefont {Zhang}}, \bibinfo {author}
  {\bibfnamefont {Z.}~\bibnamefont {Yi}}, \ and\ \bibinfo {author}
  {\bibfnamefont {L.~J.}\ \bibnamefont {Pan}},\ }\href@noop {} {\bibfield
  {journal} {\bibinfo  {journal} {Journal of Applied Physics}\ }\textbf
  {\bibinfo {volume} {121}},\ \bibinfo {pages} {134503} (\bibinfo {year}
  {2017})}\BibitemShut {NoStop}%
\bibitem [{\citenamefont {{ASTM Standard E112-13}}(2013)}]{standard1996e112}%
  \BibitemOpen
  \bibfield  {author} {\bibinfo {author} {\bibnamefont {{ASTM Standard
  E112-13}}}\ }(\bibinfo  {publisher} {ASTM International},\ \bibinfo {address}
  {West Conshohocken, PA},\ \bibinfo {year} {2013})\BibitemShut {NoStop}%
\bibitem [{\citenamefont {Jacoboni}(2010)}]{jacoboni2010theory}%
  \BibitemOpen
  \bibfield  {author} {\bibinfo {author} {\bibfnamefont {C.}~\bibnamefont
  {Jacoboni}},\ }\href@noop {} {\emph {\bibinfo {title} {{Theory of Electron
  Transport in Semiconductors: A Pathway from Elementary Physics to
  Nonequilibrium Green Functions}}}},\ Vol.\ \bibinfo {volume} {165}\ (\bibinfo
   {publisher} {Springer Science \& Business Media},\ \bibinfo {year}
  {2010})\BibitemShut {NoStop}%
\bibitem [{\citenamefont {Mahan}(2013)}]{mahan2013many}%
  \BibitemOpen
  \bibfield  {author} {\bibinfo {author} {\bibfnamefont {G.~D.}\ \bibnamefont
  {Mahan}},\ }\href@noop {} {\emph {\bibinfo {title} {Many-particle physics}}}\
  (\bibinfo  {publisher} {Springer Science \& Business Media},\ \bibinfo {year}
  {2013})\BibitemShut {NoStop}%
\bibitem [{\citenamefont {Moors}\ \emph {et~al.}(2017)\citenamefont {Moors},
  \citenamefont {Sor{\'e}e},\ and\ \citenamefont
  {Magnus}}]{moors2017resistivity}%
  \BibitemOpen
  \bibfield  {author} {\bibinfo {author} {\bibfnamefont {K.}~\bibnamefont
  {Moors}}, \bibinfo {author} {\bibfnamefont {B.}~\bibnamefont {Sor{\'e}e}}, \
  and\ \bibinfo {author} {\bibfnamefont {W.}~\bibnamefont {Magnus}},\
  }\href@noop {} {\bibfield  {journal} {\bibinfo  {journal} {Microelectronic
  Engineering}\ }\textbf {\bibinfo {volume} {167}},\ \bibinfo {pages} {37}
  (\bibinfo {year} {2017})}\BibitemShut {NoStop}%
\bibitem [{\citenamefont {Giannozzi}\ \emph {et~al.}(2009)\citenamefont
  {Giannozzi}, \citenamefont {Baroni}, \citenamefont {Bonini}, \citenamefont
  {Calandra}, \citenamefont {Car}, \citenamefont {Cavazzoni}, \citenamefont
  {Ceresoli}, \citenamefont {Chiarotti}, \citenamefont {Cococcioni},
  \citenamefont {Dabo}, \citenamefont {{Dal Corso}}, \citenamefont
  {de~Gironcoli}, \citenamefont {Fabris}, \citenamefont {Fratesi},
  \citenamefont {Gebauer}, \citenamefont {Gerstmann}, \citenamefont
  {Gougoussis}, \citenamefont {Kokalj}, \citenamefont {Lazzeri}, \citenamefont
  {Martin-Samos}, \citenamefont {Marzari}, \citenamefont {Mauri}, \citenamefont
  {Mazzarello}, \citenamefont {Paolini}, \citenamefont {Pasquarello},
  \citenamefont {Paulatto}, \citenamefont {Sbraccia}, \citenamefont {Scandolo},
  \citenamefont {Sclauzero}, \citenamefont {Seitsonen}, \citenamefont
  {Smogunov}, \citenamefont {Umari},\ and\ \citenamefont
  {Wentzcovitch}}]{Giannozzi2009}%
  \BibitemOpen
  \bibfield  {author} {\bibinfo {author} {\bibfnamefont {P.}~\bibnamefont
  {Giannozzi}}, \bibinfo {author} {\bibfnamefont {S.}~\bibnamefont {Baroni}},
  \bibinfo {author} {\bibfnamefont {N.}~\bibnamefont {Bonini}}, \bibinfo
  {author} {\bibfnamefont {M.}~\bibnamefont {Calandra}}, \bibinfo {author}
  {\bibfnamefont {R.}~\bibnamefont {Car}}, \bibinfo {author} {\bibfnamefont
  {C.}~\bibnamefont {Cavazzoni}}, \bibinfo {author} {\bibfnamefont
  {D.}~\bibnamefont {Ceresoli}}, \bibinfo {author} {\bibfnamefont {G.~L.}\
  \bibnamefont {Chiarotti}}, \bibinfo {author} {\bibfnamefont {M.}~\bibnamefont
  {Cococcioni}}, \bibinfo {author} {\bibfnamefont {I.}~\bibnamefont {Dabo}},
  \bibinfo {author} {\bibfnamefont {A.}~\bibnamefont {{Dal Corso}}}, \bibinfo
  {author} {\bibfnamefont {S.}~\bibnamefont {de~Gironcoli}}, \bibinfo {author}
  {\bibfnamefont {S.}~\bibnamefont {Fabris}}, \bibinfo {author} {\bibfnamefont
  {G.}~\bibnamefont {Fratesi}}, \bibinfo {author} {\bibfnamefont
  {R.}~\bibnamefont {Gebauer}}, \bibinfo {author} {\bibfnamefont
  {U.}~\bibnamefont {Gerstmann}}, \bibinfo {author} {\bibfnamefont
  {C.}~\bibnamefont {Gougoussis}}, \bibinfo {author} {\bibfnamefont
  {A.}~\bibnamefont {Kokalj}}, \bibinfo {author} {\bibfnamefont
  {M.}~\bibnamefont {Lazzeri}}, \bibinfo {author} {\bibfnamefont
  {L.}~\bibnamefont {Martin-Samos}}, \bibinfo {author} {\bibfnamefont
  {N.}~\bibnamefont {Marzari}}, \bibinfo {author} {\bibfnamefont
  {F.}~\bibnamefont {Mauri}}, \bibinfo {author} {\bibfnamefont
  {R.}~\bibnamefont {Mazzarello}}, \bibinfo {author} {\bibfnamefont
  {S.}~\bibnamefont {Paolini}}, \bibinfo {author} {\bibfnamefont
  {A.}~\bibnamefont {Pasquarello}}, \bibinfo {author} {\bibfnamefont
  {L.}~\bibnamefont {Paulatto}}, \bibinfo {author} {\bibfnamefont
  {C.}~\bibnamefont {Sbraccia}}, \bibinfo {author} {\bibfnamefont
  {S.}~\bibnamefont {Scandolo}}, \bibinfo {author} {\bibfnamefont
  {G.}~\bibnamefont {Sclauzero}}, \bibinfo {author} {\bibfnamefont {A.~P.}\
  \bibnamefont {Seitsonen}}, \bibinfo {author} {\bibfnamefont {A.}~\bibnamefont
  {Smogunov}}, \bibinfo {author} {\bibfnamefont {P.}~\bibnamefont {Umari}}, \
  and\ \bibinfo {author} {\bibfnamefont {R.~M.}\ \bibnamefont {Wentzcovitch}},\
  }\href {\doibase 10.1088/0953-8984/21/39/395502} {\bibfield  {journal}
  {\bibinfo  {journal} {Journal of Physics: Condensed Matter}\ }\textbf
  {\bibinfo {volume} {21}},\ \bibinfo {pages} {395502} (\bibinfo {year}
  {2009})}\BibitemShut {NoStop}%
\bibitem [{\citenamefont {Bl{\"{o}}chl}(1994)}]{Blochl1994}%
  \BibitemOpen
  \bibfield  {author} {\bibinfo {author} {\bibfnamefont {P.~E.}\ \bibnamefont
  {Bl{\"{o}}chl}},\ }\href {\doibase 10.1103/PhysRevB.50.17953} {\bibfield
  {journal} {\bibinfo  {journal} {Physical Review B}\ }\textbf {\bibinfo
  {volume} {50}},\ \bibinfo {pages} {17953} (\bibinfo {year}
  {1994})}\BibitemShut {NoStop}%
\bibitem [{\citenamefont {Perdew}\ \emph {et~al.}(1996)\citenamefont {Perdew},
  \citenamefont {Burke},\ and\ \citenamefont {Ernzerhof}}]{Perdew1996}%
  \BibitemOpen
  \bibfield  {author} {\bibinfo {author} {\bibfnamefont {J.~P.}\ \bibnamefont
  {Perdew}}, \bibinfo {author} {\bibfnamefont {K.}~\bibnamefont {Burke}}, \
  and\ \bibinfo {author} {\bibfnamefont {M.}~\bibnamefont {Ernzerhof}},\ }\href
  {\doibase 10.1103/PhysRevLett.77.3865} {\bibfield  {journal} {\bibinfo
  {journal} {Physical Review Letters}\ }\textbf {\bibinfo {volume} {77}},\
  \bibinfo {pages} {3865} (\bibinfo {year} {1996})}\BibitemShut {NoStop}%
\bibitem [{\citenamefont {Bass}\ \emph {et~al.}(1985)\citenamefont {Bass},
  \citenamefont {Dugdale}, \citenamefont {Foiles},\ and\ \citenamefont
  {Myers}}]{bass1982landolt}%
  \BibitemOpen
  \bibfield  {author} {\bibinfo {author} {\bibfnamefont {J.}~\bibnamefont
  {Bass}}, \bibinfo {author} {\bibfnamefont {J.}~\bibnamefont {Dugdale}},
  \bibinfo {author} {\bibfnamefont {C.}~\bibnamefont {Foiles}}, \ and\ \bibinfo
  {author} {\bibfnamefont {A.}~\bibnamefont {Myers}},\ }\href@noop {} {\emph
  {\bibinfo {title} {{Metals: Electronic Transport Phenomena: Electrical
  Resistivity, Thermoelectrical Power and Optical Properties}}}},\ Vol.\
  \bibinfo {volume} {Landolt-B\"ornstein III 15B}\ (\bibinfo  {publisher}
  {Springer-Verlag Berlin Heidelberg},\ \bibinfo {year} {1985})\BibitemShut
  {NoStop}%
\bibitem [{\citenamefont {Popovici}\ \emph {et~al.}(2017)\citenamefont
  {Popovici}, \citenamefont {Groven}, \citenamefont {Marcoen}, \citenamefont
  {Phung}, \citenamefont {Dutta}, \citenamefont {Swerts}, \citenamefont
  {Meersschaut}, \citenamefont {van~den Berg}, \citenamefont {Franquet},
  \citenamefont {Moussa}, \citenamefont {Vanstreels}, \citenamefont {Lagrain},
  \citenamefont {Bender}, \citenamefont {Jurczak}, \citenamefont {{Van
  Elshocht}}, \citenamefont {Delabie},\ and\ \citenamefont
  {Adelmann}}]{Popovici2017}%
  \BibitemOpen
  \bibfield  {author} {\bibinfo {author} {\bibfnamefont {M.}~\bibnamefont
  {Popovici}}, \bibinfo {author} {\bibfnamefont {B.}~\bibnamefont {Groven}},
  \bibinfo {author} {\bibfnamefont {K.}~\bibnamefont {Marcoen}}, \bibinfo
  {author} {\bibfnamefont {Q.~M.}\ \bibnamefont {Phung}}, \bibinfo {author}
  {\bibfnamefont {S.}~\bibnamefont {Dutta}}, \bibinfo {author} {\bibfnamefont
  {J.}~\bibnamefont {Swerts}}, \bibinfo {author} {\bibfnamefont
  {J.}~\bibnamefont {Meersschaut}}, \bibinfo {author} {\bibfnamefont {J.~A.}\
  \bibnamefont {van~den Berg}}, \bibinfo {author} {\bibfnamefont
  {A.}~\bibnamefont {Franquet}}, \bibinfo {author} {\bibfnamefont
  {A.}~\bibnamefont {Moussa}}, \bibinfo {author} {\bibfnamefont
  {K.}~\bibnamefont {Vanstreels}}, \bibinfo {author} {\bibfnamefont
  {P.}~\bibnamefont {Lagrain}}, \bibinfo {author} {\bibfnamefont
  {H.}~\bibnamefont {Bender}}, \bibinfo {author} {\bibfnamefont
  {M.}~\bibnamefont {Jurczak}}, \bibinfo {author} {\bibfnamefont
  {S.}~\bibnamefont {{Van Elshocht}}}, \bibinfo {author} {\bibfnamefont
  {A.}~\bibnamefont {Delabie}}, \ and\ \bibinfo {author} {\bibfnamefont
  {C.}~\bibnamefont {Adelmann}},\ }\href {\doibase
  10.1021/acs.chemmater.6b05437} {\bibfield  {journal} {\bibinfo  {journal}
  {Chemistry of Materials}\ }\textbf {\bibinfo {volume} {29}},\ \bibinfo
  {pages} {4654} (\bibinfo {year} {2017})}\BibitemShut {NoStop}%
\bibitem [{\citenamefont {Sun}\ \emph {et~al.}(2009)\citenamefont {Sun},
  \citenamefont {Yao}, \citenamefont {Warren}, \citenamefont {Barmak},
  \citenamefont {Toney}, \citenamefont {Peale},\ and\ \citenamefont
  {Coffey}}]{Sun2009}%
  \BibitemOpen
  \bibfield  {author} {\bibinfo {author} {\bibfnamefont {T.}~\bibnamefont
  {Sun}}, \bibinfo {author} {\bibfnamefont {B.}~\bibnamefont {Yao}}, \bibinfo
  {author} {\bibfnamefont {A.~P.}\ \bibnamefont {Warren}}, \bibinfo {author}
  {\bibfnamefont {K.}~\bibnamefont {Barmak}}, \bibinfo {author} {\bibfnamefont
  {M.~F.}\ \bibnamefont {Toney}}, \bibinfo {author} {\bibfnamefont {R.~E.}\
  \bibnamefont {Peale}}, \ and\ \bibinfo {author} {\bibfnamefont {K.~R.}\
  \bibnamefont {Coffey}},\ }\href@noop {} {\bibfield  {journal} {\bibinfo
  {journal} {Physical Review B}\ }\textbf {\bibinfo {volume} {79}},\ \bibinfo
  {pages} {041402} (\bibinfo {year} {2009})}\BibitemShut {NoStop}%
\bibitem [{\citenamefont {Zheng}\ \emph {et~al.}(2017)\citenamefont {Zheng},
  \citenamefont {Zhou}, \citenamefont {Engler}, \citenamefont {Chawla},
  \citenamefont {Hull},\ and\ \citenamefont {Gall}}]{Zheng2017}%
  \BibitemOpen
  \bibfield  {author} {\bibinfo {author} {\bibfnamefont {P.~Y.}\ \bibnamefont
  {Zheng}}, \bibinfo {author} {\bibfnamefont {T.}~\bibnamefont {Zhou}},
  \bibinfo {author} {\bibfnamefont {B.~J.}\ \bibnamefont {Engler}}, \bibinfo
  {author} {\bibfnamefont {J.~S.}\ \bibnamefont {Chawla}}, \bibinfo {author}
  {\bibfnamefont {R.}~\bibnamefont {Hull}}, \ and\ \bibinfo {author}
  {\bibfnamefont {D.}~\bibnamefont {Gall}},\ }\href {\doibase
  10.1063/1.4994001} {\bibfield  {journal} {\bibinfo  {journal} {Journal of
  Applied Physics}\ }\textbf {\bibinfo {volume} {122}},\ \bibinfo {pages}
  {095304} (\bibinfo {year} {2017})}\BibitemShut {NoStop}%
\bibitem [{\citenamefont {Zheng}\ and\ \citenamefont
  {Gall}(2017)}]{Zheng2017B}%
  \BibitemOpen
  \bibfield  {author} {\bibinfo {author} {\bibfnamefont {P.}~\bibnamefont
  {Zheng}}\ and\ \bibinfo {author} {\bibfnamefont {D.}~\bibnamefont {Gall}},\
  }\href@noop {} {\bibfield  {journal} {\bibinfo  {journal} {Journal of Applied
  Physics}\ }\textbf {\bibinfo {volume} {122}},\ \bibinfo {pages} {135301}
  (\bibinfo {year} {2017})}\BibitemShut {NoStop}%
\bibitem [{\citenamefont {C{\'{e}}sar}\ \emph {et~al.}(2014)\citenamefont
  {C{\'{e}}sar}, \citenamefont {Liu}, \citenamefont {Gall},\ and\ \citenamefont
  {Guo}}]{Cesar2014}%
  \BibitemOpen
  \bibfield  {author} {\bibinfo {author} {\bibfnamefont {M.}~\bibnamefont
  {C{\'{e}}sar}}, \bibinfo {author} {\bibfnamefont {D.}~\bibnamefont {Liu}},
  \bibinfo {author} {\bibfnamefont {D.}~\bibnamefont {Gall}}, \ and\ \bibinfo
  {author} {\bibfnamefont {H.}~\bibnamefont {Guo}},\ }\href@noop {} {\bibfield
  {journal} {\bibinfo  {journal} {Physical Review Applied}\ }\textbf {\bibinfo
  {volume} {2}},\ \bibinfo {pages} {044007} (\bibinfo {year}
  {2014})}\BibitemShut {NoStop}%
\bibitem [{\citenamefont {C{\'{e}}sar}\ \emph {et~al.}(2016)\citenamefont
  {C{\'{e}}sar}, \citenamefont {Gall},\ and\ \citenamefont {Guo}}]{Cesar2016}%
  \BibitemOpen
  \bibfield  {author} {\bibinfo {author} {\bibfnamefont {M.}~\bibnamefont
  {C{\'{e}}sar}}, \bibinfo {author} {\bibfnamefont {D.}~\bibnamefont {Gall}}, \
  and\ \bibinfo {author} {\bibfnamefont {H.}~\bibnamefont {Guo}},\ }\href@noop
  {} {\bibfield  {journal} {\bibinfo  {journal} {Physical Review Applied}\
  }\textbf {\bibinfo {volume} {5}},\ \bibinfo {pages} {054018} (\bibinfo {year}
  {2016})}\BibitemShut {NoStop}%
\end{thebibliography}%

\appendix
\section{Transition probability for grain boundary scattering} \label{sec:appendix1}
Fermi's golden rule prescribes a transition probability $P(\veck, \veck')$ from an initial state $\mid \! i \rangle$ with wave vector $\veck$ to a final state $\mid \! f \rangle$ with wave vector $\veck'$:
\begin{equation} \label{eq:FGR}
  P(\veck, \veck') = \frac{2 \pi}{\hbar} |\left\langle f \mid V \mid i \right\rangle |^2 \delta(E_i - E_f).
\end{equation}
The squared matrix element for the Mayadas-Shatzkes grain boundary potential yields:
\begin{equation} \label{eq:Squared_ME_GB} \begin{split}
  &\left| \left\langle f \mid V^\GB \mid i \right\rangle \right|^2 \\
  &= \delta_{\veck_\perp, \veck_\perp'} \lef \frac{S}{L} \rig^2 \sum_{n, n'} \exp \left[ \imu (k_x - k_x') (x_n - x_{n'}) \right],
\end{split} \end{equation}
with $n$ and $n'$ ranging from 1 to $N$. As a result we obtain
\begin{equation} \label{eq:FGR_GB} \begin{split}
  P(\veck, \veck') &= \frac{m_x S^2}{\hbar^3 L |k_x|} \,
    \delta_{\veck_\perp, \veck_\perp'} \delta_{k_x, -k_x'} \\
  &\quad \times \sum_{n, n'} \exp \left[ \imu (k_x - k_x') (x_n - x_{n'}) \right],
\end{split} \end{equation}
where we rewrote the Dirac delta function as $\delta(E_i - E_f) = m_x L / (2 \pi \hbar^2 |k_x'|) \, \delta_{k_x, -k_x'}$.
The average of Eq.~(\ref{eq:FGR_GB}) with the Gaussian distribution function of Eq.~(\ref{eq:Gaussian}) yields the result of Eq.~(\ref{eq:GB_Scat}), with $m_x = m_\e$.

\section{Effective mass fitting for textured and untextured thin films}
\label{sec:appendix2}
\begin{figure}[tb]
  \centering
  \subfigure[\ ]{\includegraphics[width=0.9\linewidth]{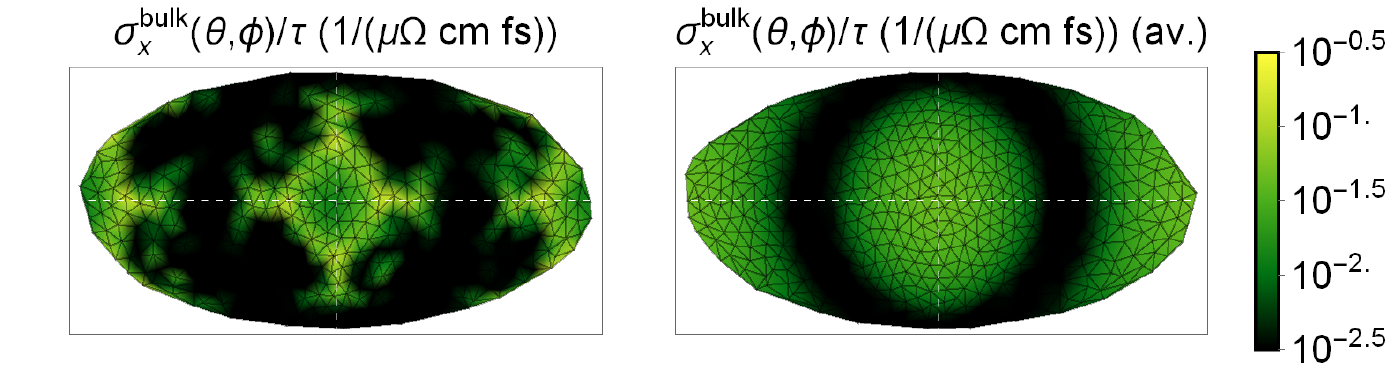}}
  \subfigure[\ ]{\includegraphics[width=0.9\linewidth]{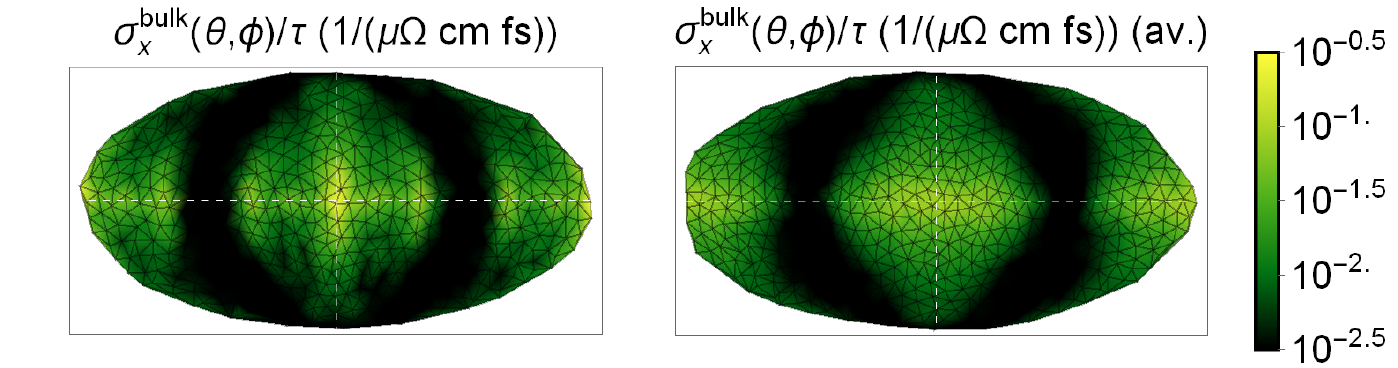}}
  \caption{
    The unaveraged (left) and averaged (right) values of $\sigma_x^\bulk(\theta, \phi)/\tau$ are shown for (a) untextured Cu with isotropic averaging, according to Eq.~(\ref{eq:averaging_isotropic}) (b) [001]-textured Ru with in-plane averaging, according to Eq.~(\ref{eq:averaging_in-plane}), considering an isotropic collision time.
    All the directions are depicted as the surface of a unit sphere in Mollweide projection with the equator and center corresponding to $z=0$ ($\theta = \pi/2$) and $z = y = 0$ ($\theta = \pi/2$, $\phi = 0$), respectively.
  }
  \label{fig:fits}
\end{figure}

\begin{table}[!tb]
  \begin{tabular}{ l c c c c }
    \hline \hline
     & & Isotropic & \\
     & Transport & quantity & $M_{x,y}^2/M_z^2$ & $\delta$ \\
    \hline
    Fit & $v_x$ & $\tau$ & 0.71 & 0.6 \\
     & & $l_0$ & 0.69 & 0.6 \\
     & $v_z$ & $\tau$ & 2.00 & 1.2 \\
     & & $l_0$ & 2.51 & 1.2 \\
     & $v_t$ & $\tau$ & 1.14 & 1.3 \\
     & ($\theta=\phi=\pi/4$) & $l_0$ & 1.31 & 1.5 \\
    \hline \hline
     & Temperature (K) & & $\rho_{x\text{-}y}/\rho_z$ & \\
    \hline
    Exp. \cite{bass1982landolt} & 100 & & 1.33 & \\
     & 200 & & 1.32 & \\
     & 300 & & 1.31 & \\
     & 400 & & 1.29 & \\
    \hline \hline
  \end{tabular}
  \caption{
    The ratios of in-plane ($\rho_{x\text{-}y}$) versus out-of-plane ($\rho_z$) resistivity of Ru are listed (represented by $M_{x,y}^2/M_z^2$) for effective mass fits with different considerations of the transport velocity direction in Eq.~(\ref{eq:cond_bulk_abinit}), assuming an isotropic collision time or mean free path.
    The normalized standard deviation $\delta$, as defined in Eq.~(\ref{eq:EM_error}), is given for each fit.
    Experimental values of the resistivity ratio are also presented for different temperatures.
  }
  \label{table:res_ratio}
\end{table}

In order to fit a diagonal effective mass tensor to a Fermi surface as presented in Fig.~\ref{fig:bands} while properly reflecting the symmetries of a textured or untextured polycrystalline thin film with differently oriented grains, an averaging procedure is introduced. We take the appropriate average over all possible orientations of the Fermi surface which can occur in the different grains. For untextured thin films, the values of $\sigma_x^\bulk(\theta,\phi)$ should be replaced with values that are obtained from averaging the \textit{ab initio} data over all angles:
\begin{equation} \label{eq:averaging_isotropic}
  \sigma_x^\bulk(\theta,\phi) \rightarrow \sin^2\!\theta \cos^2\!\phi \, \langle \sigma_t^\bulk (\theta_t, \phi_t) \rangle_{(\theta_t, \phi_t) \in \Omega}.
\end{equation}
For [001]-textured (along $z$) films, the Fermi surface should be averaged over all directions in the $x$-$y$ plane, requiring the following replacement:
\begin{equation} \label{eq:averaging_in-plane}
  \sigma_x^\bulk(\theta,\phi) \rightarrow \cos^2\!\phi \, \langle \sigma_t^\bulk (\theta, \phi_t) \rangle_{\phi_t \in [0, 2 \pi[}.
\end{equation}
Examples of this averaging procedures are shown in Fig.~\ref{fig:fits}.
Even for Cu, the directional bulk conductivity before and after averaging is very different. Nonetheless, the effective mass fitting results for Cu were found to be independent of the averaging procedure, only depending on the assumption for bulk scattering (isotropy of collision time or mean free path), as expected for a nearly isotropic Fermi surface. For Ru however, there is a strong dependence on the averaging procedure.

Up to this point, we have always considered matching the anisotropic bulk conductivity along the transport direction $x$, but one can also consider the out-of-plane bulk conductivity $\sigma_z^\bulk$ or $\sigma_t^\bulk$ along any transport direction $t$ by replacing the velocity squared that appears in the conductivity formula, $v_{x \, n}^2 \rightarrow v_{t \, n}^2$, in the derivations above and in section~\ref{subsec:fitting_procedure}.
Ideally, the resulting effective masses and Fermi energy should be consistent, but this does not appear to be the case for Ru. Its nonellipsoidal multi-band Fermi surface lies at the heart of this inconsistency. The Fermi velocities of the complicated Fermi surface get projected to the velocity squared along the direction under consideration, a process which in general does not retain all the transport features, particularly in case of highly nonellipsoidal (multi-band) Fermi surfaces.

We have fitted the effective masses for Ru based on a fitting procedure with transport along different transport directions, using the \textit{ab initio} data presented in Fig.~\ref{fig:bands}, and the results are summarized in Table~\ref{table:res_ratio}.
The ratio of in-plane versus out-of-plane resistivity, being an essential property of conduction band anisotropy and proportional to $M_{x,y}^2/M_z^2$ in the effective mass model, is strongly dependent on the transport direction under consideration.
When considering an in-plane or out-of-plane transport direction, the obtained ratios do not agree with experimental resistivity ratio. Satisfactory agreement was obtained when considering transport along the ($x=1$, $y=1$, $z=1$)-direction (or equivalently, $\theta = \phi = \pi/4$) however, with an almost perfect match when assuming bulk scattering with isotropic mean free path. We suspect that a projection of the Fermi surface velocities on this transport direction optimally retains the essential transport properties of the complicated Ru Fermi surface. We have therefore adopted the effective mass fit with the consideration of $v_t^2$ ($\theta = \phi = \pi/4$) for the comparison with experimental data in section~\ref{sec:results}.
The normalized standard deviation $\delta$, defined as
\begin{equation} \label{eq:EM_error}
 \delta \equiv \frac{1}{\sigma_x^\bulk} \sqrt{\frac{\sum_{i = 1}^N [\sigma_x^\EMA(\theta_i, \phi_i) - \sigma_x^\bulk(\theta_i, \phi_i)]^2}{N - 1}},
\end{equation}
was evaluated for each of these fits. These values are significant and one can therefore not expect the in-plane averaged directional conductivity that originates from \textit{ab initio} data to agree quantitatively with that of an ellipsoidal energy-momentum relation for arbitrary angles.

\end{document}